\documentclass[aps,pra,floatfix,twocolumn,nofootinbib,showpacs]{revtex4-1}

\usepackage{graphicx}
\usepackage{amssymb,amsmath,amsfonts,mathrsfs,fancyhdr}
\usepackage{braket}

\usepackage{epsf}
\usepackage{color}

\def\sH{\mathscr{H}}
\def\Tr{\operatorname{Tr}}

\definecolor{darkgreen}{RGB}{0, 150, 0}

\definecolor{mygray}{RGB}{150, 150, 150}

\begin{document}
\title{Key graph properties affecting transport efficiency of flip-flop Grover percolated quantum walks}
\author{J. Mare\v s, J. Novotn\'y, M. \v Stefa\v n\' ak, I. Jex}
\affiliation{Department of Physics, Faculty of Nuclear Sciences and Physical Engineering, Czech Technical University in Prague, B\v rehov\'a 7, 115 19 Praha 1 - Star\'e M\v esto, Czech Republic}
\date{\today}

\begin{abstract}
Quantum walks exhibit properties without classical analogues. One of those is the phenomenon of asymptotic trapping -- there can be non-zero probability of the quantum walker being localised in a finite part of the underlying graph indefinitely even though locally all directions of movement are assigned non-zero amplitudes at each step. We study quantum walks with the flip-flop shift operator and the Grover coin, where this effect has been identified previously. For the version of the walk further modified by a random dynamical disruption of the graph (percolated quantum walks) we provide a recipe for the construction of a complete basis of the subspace of trapped states allowing to determine the asymptotic probability of trapping for arbitrary finite connected simple graphs, thus significantly generalizing the previously known result restricted to planar 3-regular graphs. We show how the position of the source and sink together with the graph geometry and its modifications affect the excitation transport. This gives us a deep insight into processes where elongation or addition of dead-end subgraphs may surprisingly result in enhanced transport and we design graphs exhibiting this pronounced behavior. In some cases this even provides closed-form formulas for the asymptotic transport probability in dependence on some structure parameters of the graphs.
\end{abstract}

\maketitle

\section{Introduction}

Quantum walks represent a simple but versatile models exhibiting a multitude of effects resulting from quantum interference \cite{review,rev2}. Among other aspects like search \cite{skw,ambainis:search,potocek,reitzner,portugal:search,wong1,wong2,wong3} and state transfer \cite{Tanner2009,Skoupy2016}, recurrence \cite{Stefanak2008,Werner2013,bourgain,nitsche}, or topological phenomena \cite{kitagawa1,kitagawa2,Asboth2012}, the rate of spreading of quantum walks across the underlying medium \cite{Aharonov2001} (usually represented by a graph) and their hitting times \cite{Magniez2012} were studied. It was discovered that on one side quantum walks can provide a quadratic \cite{quantum_random_walks} or in special cases even exponential \cite{glued_trees} speedup in propagation, but on the other side can also exhibit various types of propagation inhibition including the Anderson localization \cite{anderson, anderson_qw,joye1,ahlbrecht1,joye2,ahlbrecht2}. Even more striking was the further discovery of trapped states \cite{Inui2004}. Trapped states are eigenstates of the dynamics with a strictly limited support in the position space of the walk - their overlap with some parts of the position space is purely zero. These states may induce permanent trapping of the walker at the vicinity of the origin of an infinite graph with non-zero probability \cite{Inui2004,inui:psa,inui:grover1}. The same mechanism allows the quantum walker to indefinitely evade an absorbing sink in a finite graph \cite{krovi1,krovi2,krovi3,theory} representing another interesting and potentially useful application. In quantum search, trapped state can cause imperfect detection \cite{thiel1,thiel2}.

Similar to other quantum phenomena, trapped eigenstates are prone to being destroyed by external disturbances of the quantum walk. With this respect a delicate intermediate position between pure unitary evolution and completely classical behavior is occupied by quantum walks with dynamical percolation, where edges of the underlying graph, representing the position space, are being made available or prohibited for the walker randomly during the time evolution \cite{asymptotic1}. This kind of external intervention or noise allows the walk to transition asymptotically into a regime with significantly reduced dynamics. Some of the trapped states are removed in the process, but other will remain present and the simplification of the asymptotic dynamics reveals them and may allow their complete classification. In addition, these surviving states offer interesting information about the original unitary dynamics.

Being the result of degeneracy of the quantum coin operator, trapped states are present in quantum walks with particular coin operators in graphs with vertices of degree 3 or higher \cite{lazy_first_mentioned,classification:1d,classification_trapping_2d}. A genuine example of a quantum walk with trapped states is the one with the flip-flop shift operator and the Grover coin. The classification of the corresponding trapped states in percolated Grover quantum walks was already carried out for simple connected planar 3-regular finite graphs \citep{theory}. The present paper extends the results to all simple connected finite graphs and identifies essential graph properties including positions of the source and the sink, whose mutual interplay gives rise to intriguing transport features. This level of generality allows to understand these effects and also to construct tailored graphs or modify a given ones with a predictable influence of the transport properties. We also provide examples of graphs where analytical approach provides closed-form formulas for the asymptotic transport probabilities.

The present paper has the following structure. In section \ref{definition} we give the necessary definitions of flip-flop Grover quantum walks both with and without percolation, introduce formally the concept of trapped states and the asymptotic transport probability, and also provide the recipe for construction of the subspace of trapped states in percolated Grover quantum walks on arbitrary simple connected graphs. The technical part of this section related to the proof of completeness of trapped subspace is presented in the necessary details in appendix \ref{appendix_non_p}. Section \ref{designing_graphs} uses results from section \ref{definition} to investigate the influence of different graph structures and choices of the source and sink placements on the transport properties. A summary and concluding remarks are left for section \ref{conclusions}. 

\section{Quantum walk definition and trapped states}
\label{definition}

We study coined (discrete time) quantum walks and in particular we employ the definition presented in \citep{theory}. In short, a quantum walk is based on an undirected graph $\mathcal{G}(V,\mathcal{E})$ called the structure graph, with the set of vertices $V$ and the set of edges $\mathcal{E}$. On top of the structure graph we define the state graph $G(V,E)$ having the same set of vertices $V$, a pair of directed edges $e_1,e_2 \in E$ going in opposite directions for every undirected edge $\epsilon \in \mathcal{E}$ with the possibility to add unpaired directed self-loops $l\in L \subset E$ as needed. See example graphs in FIG. \ref{fig:connecting_states}. We denote $\tilde{e}_1 = e_2, \tilde{e}_2 = e_1, \tilde{l} = l$ and also the support edge $|e_1|=|e_2|=\epsilon$. The Hilbert space of the walk is spanned by base states corresponding to the directed edges in the state graph as $\mathcal{H}=\mathrm{span}\{\ket{e}| e\in E\}$. We further define a vertex subspace $\mathcal{H}_v$ for every vertex $v\in V$ as $\mathcal{H}_v=\mathrm{span}\{\ket{e}| o(e) = v\}$, where $o(e)$ is the origin vertex of the edge $e$. 

The time evolution proceeds in discrete steps realized by a unitary operator $U$, which consists of the shift operator $S$ and the coin operator $C$ as
\begin{align*}
\ket{\psi(t+1)} = U\ket{\psi(t)} = SC\ket{\psi(t)}.
\end{align*}
In this work we focus on quantum walks with the flip-flop (reflecting) shift operator $S=R$, which just swaps amplitudes for pairs of edges $e$ and $\tilde{e}$. The Grover coin acts by application of a Grover matrix of the corresponding dimension $d$ in every vertex subspace:
\begin{align*}
G_d &= 2 \vert \phi_d\rangle\langle \phi_d \vert - I_d,
\end{align*}
where $\vert \phi_d\rangle =
(\vert 1\rangle + \vert 2\rangle + \ldots + \vert d\rangle )/\sqrt{d}$. The flip-flop Grover walk constitutes a natural choice as it is well defined for an arbitrary graph with vertices of different degrees. Moreover, since Grover matrices commute with all permutation matrices, the ordering of base states and associated amplitudes within vertices does not play a role and thus can be arbitrary.

Following \citep{theory}, we introduce so called dynamical percolation of the underlying graph - in every step some edges are chosen randomly\footnote{Every edge has a probability $\pi$ to be open and $1-\pi$ to be closed in every step independently of the others. The value of $\pi$ is irrelevant for the asymptotic behavior as long as it is non-zero. It only affects the rate of convergence \cite{ruo}.} to be closed (and reopen for further steps) not allowing the walker to pass. For a given configuration of open edges $K\subset E$ this results in a modification of the shift operator to $R_K$ so that the closed edges are treated as unpaired loops:
\begin{equation*}
R_K=\sum_{e\notin L, |e|\in K}\ket{\tilde{e}}\bra{e} + \sum_{e\notin L, |e|\notin K}\ket{e}\bra{e} + \sum_{l\in L}\ket{l}\bra{l}.
\end{equation*}
Due to lack of detailed control over the system, the time step of a Percolated Coined Quantum Walk (PCQW \footnote{By CQW we always mean a coined quantum walk without percolation.}) is a statistical mixture over all possible configurations described by the so called random unitary operation
\begin{equation}
\rho(t+1)= \sum_{K\subset E}\pi_KU_K \rho(t) U_K^\dagger = {\cal U}(\rho(t)),
\label{timeevolutionpercolated}
\end{equation}
where $U_K = R_K C$ and $\pi_K$ denotes the probability of the configuration $K$ being chosen. A general approach for finding the asymptotic regime of a system driven by a random unitary operation was given in \cite{ruo}. The asymptotic state is described by
\begin{equation}
\label{as_state}
\rho_{as}(t) = \sum_{\lambda,i}\lambda^t\Tr{\left(\rho(0) X_{\lambda,i}^\dagger \right)} X_{\lambda,i},
\end{equation}
where $\lim_{t\rightarrow \infty} || \rho(t) - \rho_{as}(t)|| = 0$. $X_{\lambda,i}$ are so called \textit{attractors} (eigen-matrices of the evolution operator $\cal U$ with eigenvalues $\lambda$ fulfilling $|\lambda|=1$) forming a basis of the asymptotic subspace, distinguished by the eigenvalue $\lambda$ and an index $i$ in case of a degeneracy. It is shown in \cite{ruo} that the attractors are exactly the common solutions of the following set of equations
\begin{equation}
U_K X U_K^\dagger = \lambda X\,, ~~{\rm for~all}~K\in 2^\mathcal{E},
\label{attractors_main}
\end{equation}
where $|\lambda|=1$. The search for attractors can be simplified with the concept of p-attractors introduced in \cite{asymptotic2}. The p-attractors are special attractors constructed with arbitrary coefficients $A^{\alpha, i}_{\beta, j}$ as
\begin{equation}
\label{p_attractors_construction}
Y_\lambda = \sum_{\alpha\beta^* = \lambda, i, j} A^{\alpha, i}_{\beta, j} \ket{\phi_{\alpha, i}}\bra{\phi_{\beta, j}}
\end{equation}
from so called common eigenstates - vectors simultaneously solving all equations
\begin{equation}
\label{p_attractors_equation_main}
U_K \ket{\phi} = \alpha \ket{\phi}\,, ~~{\rm for~all}~K\subset 2^\mathcal{E}.
\end{equation}
For the case of quantum walks the set of equations (\ref{p_attractors_equation_main}) can be converted to two conditions, which need to be fulfilled simultaneously and can be stated in the following form: the local coin condition 
\begin{align}
\label{coin_condition_common_eigenstaes_locally}
G_d \ket{\phi}_v &= \lambda \ket{\phi}_v\,, ~~{\rm for~all}~v\in V,
\end{align}
where $\ket{\phi}_v$ is the restriction of the state $\ket{\phi}$ to the vertex subspace $\mathcal{H}_v$ of dimension $d$, and the shift condition
\begin{align}
\label{p_shift_main}
\phi_{e}=\phi_{\tilde{e}}\,, ~~{\rm for~all}~e\in E,
\end{align}
where $\phi_{e}$ is the element of $\ket{\phi}$ corresponding to the directed edge $e$. We use these conditions to construct the basis of common eigenstates and, thereby, also the p-attractors. A detailed construction of common eigenstates follows below and for more details on derivation of the conditions (\ref{p_shift_main}) see Appendix \ref{appendix_non_p}.

The p-attractors form only a subspace in the asymptotic subspace, but in some cases only a single non-p-attractor (proportional to the identity matrix) needs to be added. It was successfully demonstrated in \citep{theory} that this is the case for the flip-flop Grover PCQWs on finite simple connected planar structure graphs with maximal vertex degree 3 and 3-regular state graphs. In the present paper we extend this result to arbitrary finite simple\footnote{The requirement of simplicity of the structure graph is mostly just for convenience. The presence of parallel edges (multiple edges connecting the same pair of vertices) and undirected self-loops in the structure graph mostly do not pose a principal problem, but make the notation and terminology more complicated.} connected structure graphs, where the degrees can vary among vertices and the graphs do not need to be planar. The absence of non-p-attractors different from the identity operator is a crucial simplification of the search for the asymptotic dynamics of a percolated quantum walk. We leave the proof for the Appendix \ref{appendix_non_p} for its length.

We aim to study transport and so we further introduce sink into the quantum walk as detailed in \citep{ladder_cayley}. We choose some vertices to act as the sink and form a corresponding sink subspace $\sH_s$ and define the projector onto the sink subspace $\Pi_{\sH_s}$ and the complement projector $\Pi = I-\Pi_{\sH_s}$. At the end of each time step there is a projection of the walker's state on the complement of the sink subspace. The state of the walker is described by a density matrix $\rho(t)$ and one step of the PCQW with sink is realized as
\begin{equation}
\label{step_sinked_percolated}
\rho(t+1)=\sum_{K\subset E}\pi_K \Pi\left( U_K \rho(t) U_K \right)\Pi\, = {\cal L}(\rho(t)).
\end{equation}
Due to the sink, the operator $\cal L$ is not trace preserving. The value $\Tr\left( \rho(t) \right)$ represents the probability of the walker still avoiding the sink. While a classical random walker on a finite connected graph inevitably reaches the sink as the time goes to infinity, the interference in quantum walks allows for existence of attractors of the dynamics without overlap with the sink. Indeed, attractors of the PCQW with sink are simply attractors of the PCQW satisfying additional condition $\Pi X \Pi=X$, i.e. attractors of $\cal U$ having no overlap with the sink. Since in our studied case this excludes the only non-trivial non-p-attractor of $\cal U$ proportional to the identity, the attractor space of our PCQW with sink consists of p-attractors only, which are constructed from common eigenstates of the map $\cal U$ with zero overlap with the sink. We call common eigenstates of $\cal U$ having limited support, i.e. having zero amplitudes for some base states, as trapped states. If the trapped state in addition has no overlap with the sink subspace, we call it a sink-resistant trapped state (sr-trapped state). 
We define the asymptotic trapping probability for the initial state $\rho_0$ as $p(\rho_0) = \lim\limits_{t \rightarrow +\infty}\Tr\left({\cal L}^{t}(\rho_0)\right)$ with ${\cal L}^t={\cal L} \circ {\cal L}^{t-1}$, and the Asymptotic Transport Probability (ATP)
\begin{equation}
\label{def_efficiency}
q(\rho_0) = 1-p(\rho_0) .
\end{equation}
Thus, when the subspace of the sr-trapped states is available with a projector to this subspace $\Pi_T$, the ATP can be calculated as
\begin{align}
\label{overlap}
q(\rho_0)=1-\Tr(\Pi_T \rho_0).
\end{align}

When we talk about transport by quantum walk, the initial state is usually chosen to be localized in one or just several vertices. We define an initial subspace, where the initial state is allowed to have non-zero elements. The influence of the initial state on the ATP is crucial. For a general consideration of a particular walk it can be beneficial to calculate the average ATP
\begin{align}
\label{overlap_average}
\bar{q}= 1- \Tr(\Pi_T \overline{\rho}),
\end{align}
where $\overline{\rho}$ is the maximally mixed state in the initial subspace.

Let us now search for the common eigenstates, which allow the construction of the projector to the subspace of sr-trapped states and the calculation of the ATP for given initial states. Note that they form a subset of eigenstates of non-percolated CQWs derived in \citep{etsuo_symmetry}, since the original structure graph represents one of the possible configurations in (\ref{p_attractors_equation_main}). However, the common eigenstates have to fulfill (\ref{p_attractors_equation_main}) for all configurations of open edges, which leads to stricter conditions (\ref{coin_condition_common_eigenstaes_locally}) and (\ref{p_shift_main}). Since the spectrum of the Grover matrix contains only two eigenvalues $1$ and $-1$, the coin condition (\ref{coin_condition_common_eigenstaes_locally}) eliminates all eigenstates of non-percolated CQW corresponding to $\lambda\neq \pm 1$. We show that for $\lambda = 1$ there is only one common eigenstate and it is not an sr-trapped state.

We start with the coin condition (\ref{coin_condition_common_eigenstaes_locally}), which implies that locally at each vertex $v$ the common eigenstate reduces to an eigenvector of the Grover matrix corresponding to the same eigenvalue (or a zero vector, but it has to be non-vanishing for some $v$). The eigenvector of the Grover matrix associated with the eigenvalue $1$ has all the elements the same
\begin{align}
\label{phi1}
\ket{\phi^1}_v =
\left[
\begin{array}{c}
 1 \\
 \vdots \\
 1 
\end{array}
\right].
\end{align}
The whole remaining subspace orthogonal to the vector $\ket{\phi^1}_v$, i.e. vectors with the sum of all elements equal to zero, is formed by eigenvectors corresponding to the eigenvalue -1. The common eigenstates are formed from the single-vertex blocks by applying the shift condition (\ref{p_shift_main}) -- the elements corresponding to every pair of directed edges on the same support edge must be the same. For the eigenvalue 1 the one vertex form (\ref{phi1}) combined with the shift condition directly implies that there is just one common eigenstate: a vector with all elements equal. Clearly, this state overlaps with the sink regardless of the position of the sink and so it is not an sr-trapped state. 

The common eigenstates corresponding to the eigenvalue $-1$ form the only part relevant for the ATP (\ref{def_efficiency}). Interestingly, as we show, the common eigenstates for the eigenvalue -1 derived below are exactly the eigenstates for the eigenvalue -1 for the non-percolated version of the walk presented in \citep{etsuo_symmetry}, so none of the states is removed by percolation. Here we reuse the key idea of utilising fundamental cycles of a graph to construct these states as in \citep{etsuo_symmetry} but we approach the construction in an alternative way.

First we determine the dimension of the subspace of common eigenstates associated with the eigenvalue -1 and then we construct the required number of linearly independent common eigenstates. The whole Hilbert space has dimension $\sum_{v\in V}d(v) = 2|E| + |L|$, where $d(v)$ is the degree of the vertex $v$ and $|\bullet|$ stands for the number of elements in the given set. The coin condition (\ref{coin_condition_common_eigenstaes_locally}) gives $|V|$ equations reducing the dimension of the subspace of common eigenstates (the sum of elements of the common eigenstate must be zero in each vertex). The shift condition gives $|E|$ equations. By directly following the derivation in \citep{theory}, one can show that also for general graphs all the equations from the coin condition and the shift condition are independent except for the case of a bipartite structure graph with a state graph without unpaired loops. In that case one condition is dependent. We, therefore, search for either $N=2|E|+|L|-|V|-|E| = |E|-|V|+|L|$ or $N+1 = |E|-|V|+1$ linearly independent common eigenstates respectively for the two cases. Note that this exactly coincides with the dimension of the -1 subspace for the non-percolated version of the walk \cite{etsuo_symmetry}.

\begin{figure}
	\centering
	\includegraphics[width=230 pt]{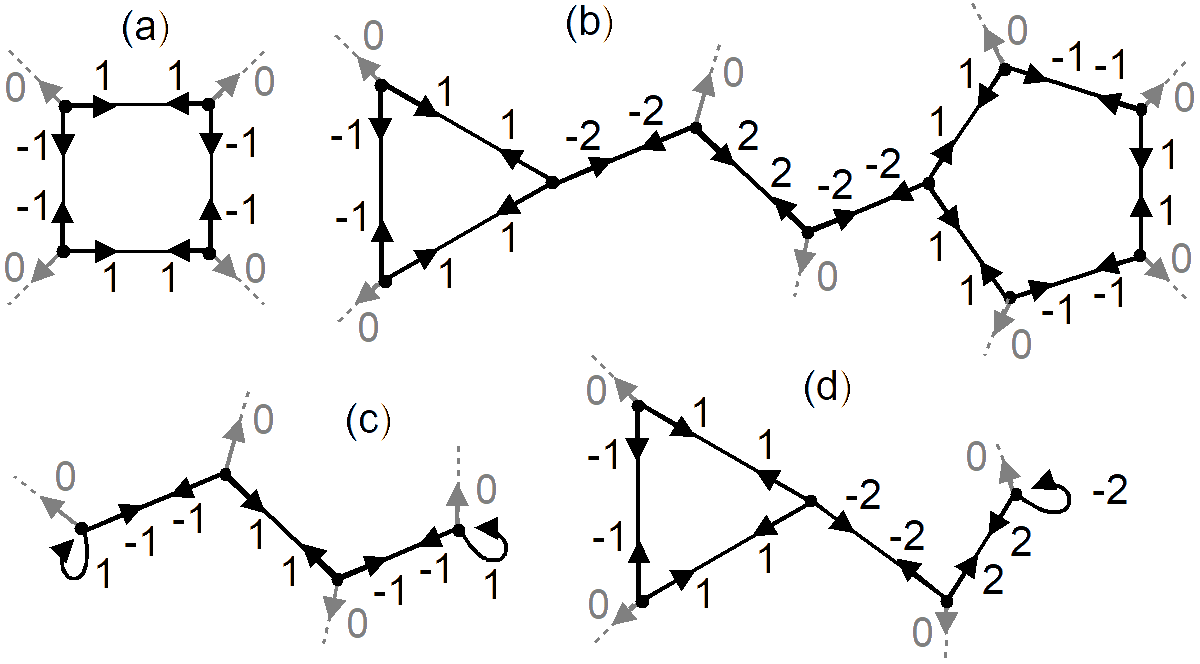}
	\caption{The four basic types of possible eigenstates for the eigenvalue -1: (a) an A-type state with alternating +1 and -1 elements on one even cycle and (b) B-type, (c) C-type and (d) D-type states each with non-zero elements on two termination elements (odd cycles or unpaired loops) and a connecting path. Dashed lines represent an arbitrary continuation of the graph where all the corresponding vector components of the given state are 0. All the states clearly follow both the coin condition and the shift condition. For a detailed construction of the states and assignment of particular elements see \citep{theory}.}
	\label{fig:trapped_construction}
\end{figure}

Let us explain in detail how these eigenstates can be systematically constructed using the fundamental cycles. From any connected graph we can construct a spanning tree -- a subgraph obtained by removing some edges which is a connected tree and still contains all the vertices from the original graph. Note that a spanning tree is not unique and a different choice may lead to a different basis of common eigenstates. If we recover any one of the removed edges, we obtain a graph with exactly one cycle -- a fundamental cycle corresponding to the recovered edge. The spanning tree has $|V|$ vertices and, as it is a tree, it has $|V|-1$ edges. Therefore, there are $|E|-|V|+1$ edges in the original graph not present in the spanning tree and we can construct the same number of fundamental cycles. The fundamental cycles and unpaired loops are used to construct common eigenstates as exemplified in Fig. \ref{fig:trapped_construction}. There are four types of common eigenstates. A-type states simply correspond to even fundamental cycles. The remaining three types consist of a connecting path on the spanning tree terminated on its both sides either by an unpaired loop or an odd fundamental cycle, which we collectively call termination elements. The magnitudes and signs of the elements in different parts of the states are easily seen in FIG. \ref{fig:trapped_construction}. Here we stress that the length of the connecting path can be zero and terminating fundamental odd cycles can even share one or more edges in the case of B-type states. In such a case, coefficients of shared edges are equal to zero and we basically obtain an A-type state \footnote{By joining two such odd cycles and eliminating the shared edges we obtain an even cycle}.

In the following we show that the above described common eigenstates associated with eigenvalue $-1$ allow us to construct basis with the required number of vectors. We denote the numbers of even and odd fundamental cycles as $|FC_e|$ and $|FC_o|$ respectively. An undirected graph is bipartite if and only if it has no odd cycles. Therefore, we need to construct $N+1=|E|-|V|+1$ states only in cases when $|FC_o|=0$ and $|L|=0$. We simply use $|FC_e|=|E|-|V|+1$ A-type states. In all other cases we need $N=|E|-|V|+|L|$ states and we have a non-zero number of termination elements. We again use $|FC_e|$ A-type states and add $|FC_o| + |L| -1$ states of other types following two rules: each new state uses a termination element not used before and the connecting paths are restricted to the spanning tree. In total, we have $|FC_e|+|FC_o| + |L| -1=|E|-|V|+1+|L|-1=N$ states as needed. In all cases, linear independence of our set of states is trivially seen as every state has some non-zero elements, which have zero values in all the others. It is either on the recovered edges in fundamental cycles or on the unpaired loops. 

This construction is suitable for any graph in question and provides a basis of the trapped states by a straightforward algorithm when a spanning tree of the graph is chosen. Nevertheless, in some cases a different approach is more convenient. When a state is added into the basis, we can use the edge recovered for this state in the construction of the other states instead of only using the edges on the spanning tree. This is equivalent to replacing the new state constructed on the recovered edge and the spanning tree by its linear combination with one or more states already present in the basis. This is in particularly useful in graphs with many adjacent even cycles as will be demonstrated on prism graphs below.

Until now, we did not consider the role of the sink, so we only searched for the trapped states. In sr-trapped states, all the amplitudes in the sink subspace must be zero. The shift condition also forces the other edges in pairs with sink edges to be zero. Otherwise, we are in the same situation as before. We can again use an approach from \citep{etsuo_symmetry}. We simply repeat the above construction, now on a reduced structure graph $\mathcal{G}_0=(V_0, \mathcal{E}_0)$ with the sink vertices and their adjacent edges removed and a state graph $G_0=(V_0, E_0)$ reduced accordingly. The graphs may loose connectedness, so we just apply the approach on each component separately. In this way we construct a maximal subspace of the trapped states subspace which is orthogonal to the sink subspace.

While the constructed states form a basis of the asymptotic subspace, they are not mutually orthogonal in general. Unfortunately, there is no general procedure available for their analytical orthogonalization and numerics or special approaches in particular cases \citep{ladder_cayley,etsuo_ladder} must be used. Nevertheless, knowledge of all trapped states provides a deep insight allowing to construct graphs for quantum walks exhibiting interesting transport properties. Numerical orthogonalization of the trapped states can then be used easily for particular graphs.

\section{Key factors for the transport efficiency and examples}
\label{designing_graphs}

Equipped with the structure of trapped states of the flip-flop Grover percolated quantum walk for any chosen graph we can now discuss the role of individual ingredients in the transport efficiency. The ATP is given as the overlap of the sr-trapped states subspace with the initial state according to (\ref{overlap}). The structure graph and the added unpaired loops define the set of trapped states, the placement of the sink selects a subset of sr-trapped states, and the choice of the initial subspace defines the scope of possible initial states with certain overlap with the sr-trapped states. In the following, we investigate the influence of the above choices and their mutual interplay on the resulting ATP. 

\subsection{The initial subspace}
The initial state plays a crucial role in transport. In particular, if we choose the initial state with the same elements for all outgoing edges in one vertex, it is always fully transported as it is by construction orthogonal to all trapped states. As the other extreme, if the initial subspace can accommodate an entire sr-trapped state (e.g. an initial vertex with two unpaired loops), choosing it as the initial state leads to complete trapping. 

\begin{figure}
  \includegraphics[width=0.18\textwidth]{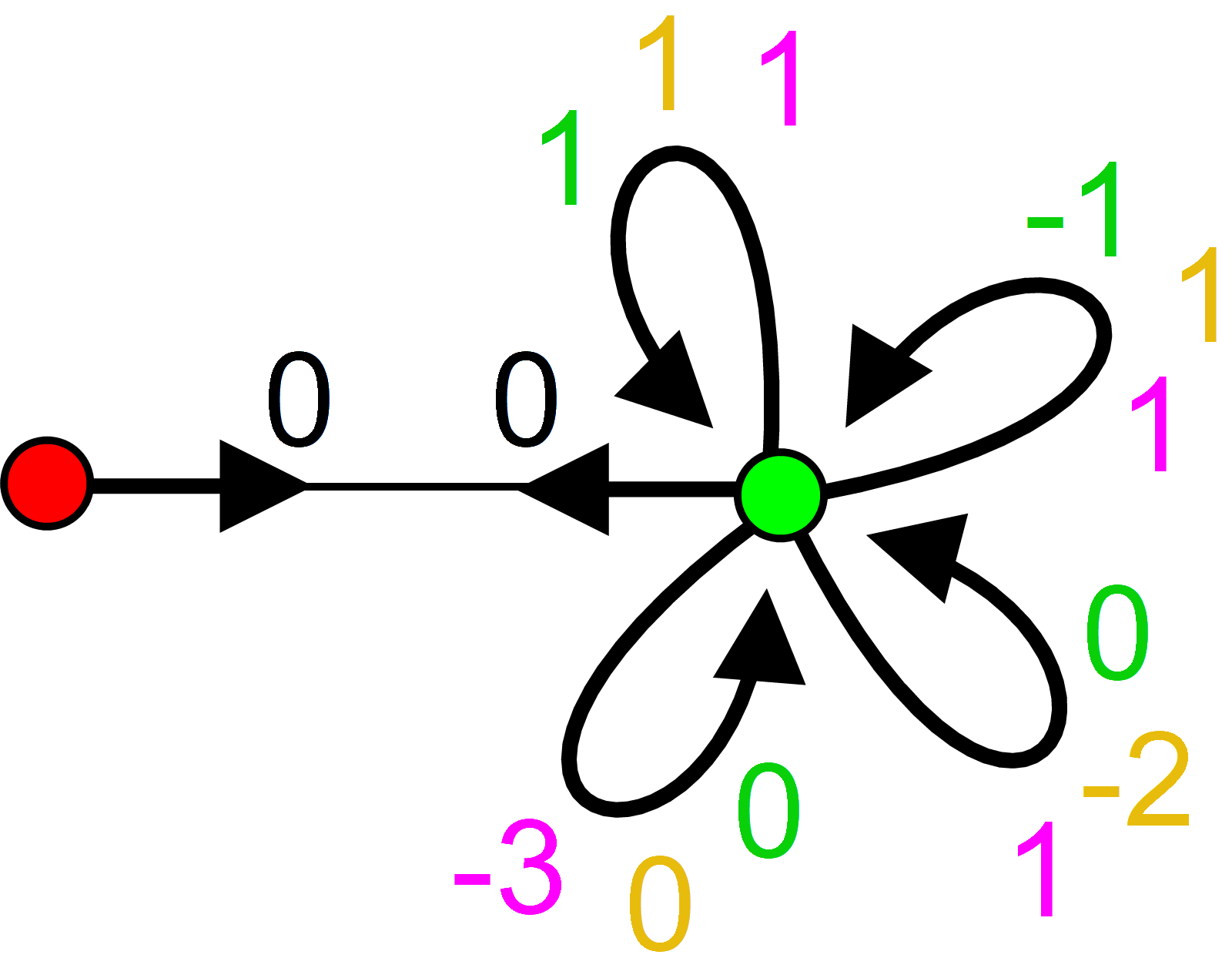}
  \caption{An example graph with multiple unpaired loops added in the initial vertex. The sink is depicted in red and the initial vertex in green. Three different orthogonalized trapped states are shown with elements of different colors.}
  \label{fig:multiple_loops}
\end{figure}

Let us now explore the extreme cases when focusing on the average ATP given by (\ref{overlap_average}). A state graph ensuring complete transport case is again simple to design. Whenever $\mathcal{G}_0$ has no even cycles and there is at most one termination element, the walker is always fully transported as the graph does not admit any sr-trapped states. The opposite task is in fact impossible. The common eigenstate for the eigenvalue 1 will always overlap with the initial subspace and will, therefore, have a non-zero contribution to the ATP. To minimize the average ATP, an effective strategy is to add unpaired loops into the initial subspace. This is seen in \citep{ladder_cayley} where by removing a dead-end branch of a Cayley tree, where the walker could seemingly be trapped, the average ATP actually decreases. That is because the branch is replaced by a loop in the initial vertex, which causes more trapping than the trapped states supported on the whole branch. For the simplest example of a graph with only two connected vertices where one acts as the sink and the other one represents the initial subspace as shown in FIG. \ref{fig:multiple_loops} we can even derive a closed-form expression for the ATP in dependence on an increasing number of added unpaired loops in the initial vertex. When the degree of the initial vertex is $n\geq 2$ there are $n-1$ unpaired loops and we can simply form an orthogonal basis of the subspace of sr-trapped states from $n-2$ C-type states. An orthogonal basis can be formed so that the $k-$th state has the element 1 on the first $k$ loops and the element $-k$ the the next one. In fact, for the calculation of the ATP we would not even need to know the orthogonal form of the states. As they are entirely contained inside the initial subspace, the overlap of each of them with the maximally mixed state on the initial subspace is simply $1/n$. Therefore, the average ATP is
\begin{align}
\label{loops_exampe}
\overline{q} = 1 - (n-2)\frac{1}{n} = \frac{2}{n},
\end{align}
which approaches zero in the limit, but it is always positive. 

\subsection{No trapping along the way}

\begin{figure}
  \includegraphics[width=0.20\textwidth]{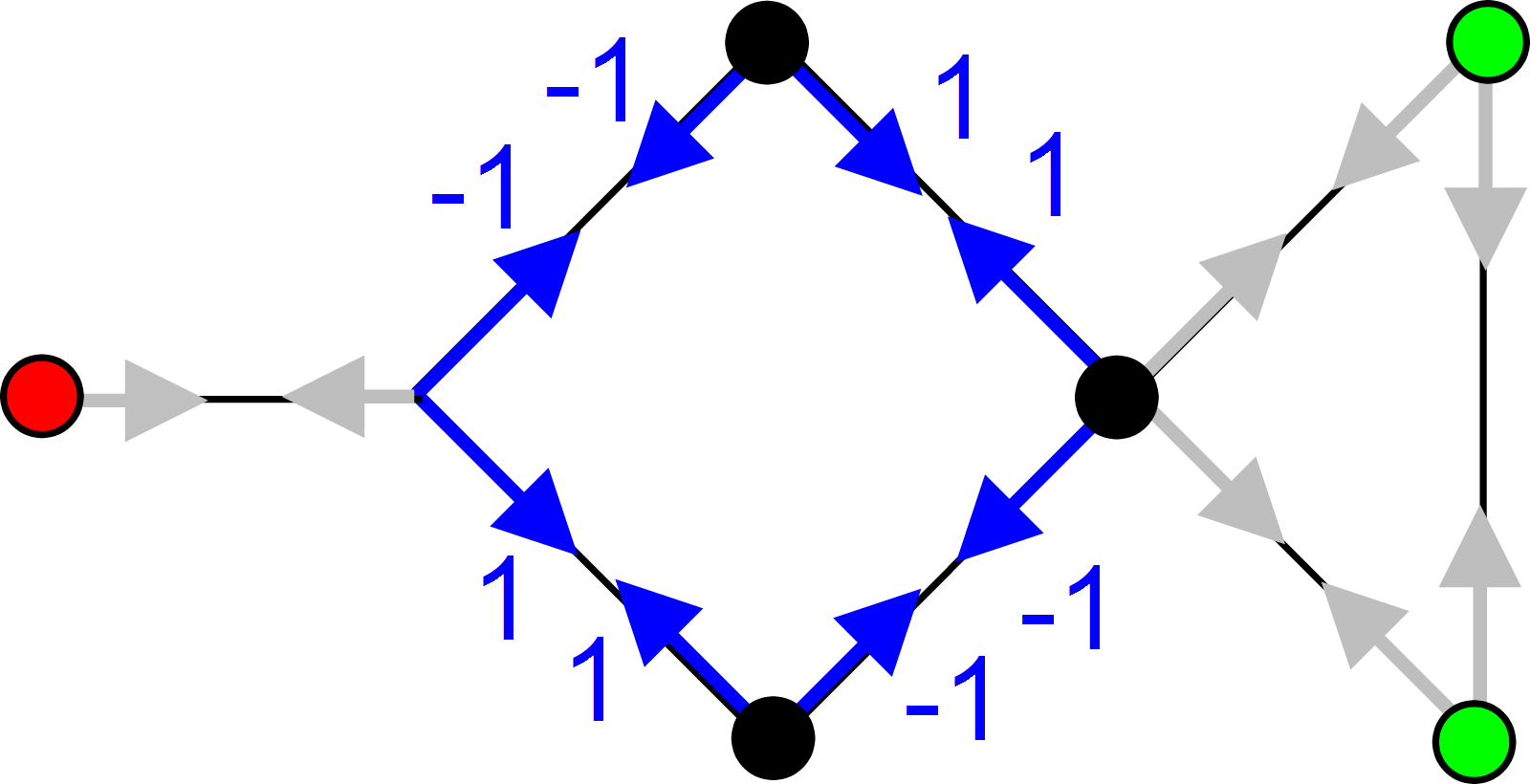}
  \caption{An example graph, where the initial subspace (green vertices) has no overlap with the only sr-trapped state present. While the walker must cross the the inner square accommodating the trapped state to reach the sink (the red vertex), there is zero chance of trapping the walker along the way. Note that if the square is replaced by an odd cycle, the situation changes. Indeed, from the two odd cycles we form a B-type sr-trapped state partially supported on the initial subspace.}
  \label{fig:no_trapping}
\end{figure}

When considered in terms of trapped states, it is a trivial observation that there is no trapping if the sr-trapped subspace does not overlap with the initial state. Nevertheless, in a classical view this may seem rather counter-intuitive that the walker can not be trapped even in situations where it is necessary to pass through parts of the graph accommodating trapped states to reach the sink. This situation is shown in FIG. \ref{fig:no_trapping}. The graph has one odd and one even cycle, which results in a single A-type trapped state. This state does not overlap with the initial state, so the ATP is equal to 1 for any initial state of the 4-dimensional initial subspace despite the fact that the walker must traverse the inner square to reach the sink. 

There is actually a reasoning for this behavior based on the local properties of the walk. We call the vertices and edges in fundamental cycles and on a possible connecting path for some trapped state the support of this state. If the initial state does not lie on the support of a trapped state, the walker must enter the support by a crossing as in the vertex between the square and the triangle in FIG. \ref{fig:no_trapping}. When any amplitude enters a crossing vertex, the Grover coin assigns the same resulting amplitudes to all outgoing edges except the direction of the incoming walker. Therefore, the two directions on the support of a trapped state always receive exactly the same amplitudes. As we know, such final state is always orthogonal to the trapped state in question. Thanks to linearity, this holds even if multiple different amplitudes are incoming into the crossing vertex simultaneously.    

It is good to note here that trapped states originally not overlapping with the initial subspace can contribute to trapping via overlap with other trapped states that do overlap with the initial subspace. Therefore, these need to be considered in orthogonalization of the basis of sr-trapped states.

\subsection{Geometry modified by pure addition}

Let us consider modifications of the graph geometry outside of the initial subspace, i.e. the initial subspace is now left unchanged. We distinguish two basic types. The first one is \textit{pure addition} of vertices and edges, both undirected in the structure graph or unpaired loops in the state graph. These are changes which keep all previous vertices with their links and simply add new ones including unpaired loops. Apparently, pure addition can never increase the ATP or, in other words, it never decreases trapping. This is a simple consequence of the fact that all the sr-trapped states present before the addition are also present after it and the overlap of the original trapped states with the initial subspace is unchanged. This is true even if we are adding new sink vertices. Here we clearly mean truly adding new vertices. If we just broaden the sink subspace to some vertices of the present graph, the ATP can increase.

We can illustrate the effect of pure addition on a star-like graph. Consider a root vertex of degree $n\geq 3$ accommodating the initial subspace and $n$ branches - chains of the same length $L$ attached to the root each being formed by $L\geq 1$ vertices of degree 2 in the state graph with the last one on each branch having one unpaired loop. Without the loops, no trapping would be present. One of the chains is terminated by a sink vertex. This example can be thought of as an extension of the one presented in FIG. \ref{fig:multiple_loops}, but for demonstration of pure addition we need to keep the initial subspace independent of the number of branches. Only a single fixed directed edge in the root (not on the sink branch) acts as the initial subspace. There are $n-1$ unpaired loops on the branches not containing the sink giving rise to $n-2$ C-type states. An orthogonal basis of the subspace of sr-trapped states can also be constructed as follows: For each base state the elements in the root vertex are the same as in  FIG. \ref{fig:multiple_loops} and we just extend the C-type states on the chains (same elements on unpaired loops and alternating signs in vertices). The $k-$th state is orthogonal to the $m-$th ($m<k$) since the contributions to the scalar product from the first $m-1$ branches are exactly compensated by the contribution from the $m-$th branch. The normalization constant for the $k-$th state is $N_k = \sqrt{(2L+1)(k+k^2)}$. Then the average transport probability is  
\begin{align}
\label{star_single_initial}
\overline{q} = 1 - \sum_{k=1}^{n-2}\frac{1}{N_k^2}\frac{1}{1} = 1 - \frac{1}{2L+1}\left(1-\frac{1}{n-1}\right).
\end{align}
Looking at (\ref{star_single_initial}) we can see that by adding new branches (increasing $n$) the average ATP decreases as predicted. In contrast, lengthening of the branches (increasing $L$) increases the ATP. This is possible since lengthening of branches is no more a pure addition but insertion as will be discussed in the next section. 

We can also consider a situation where the initial subspace is extended with every added branch to cover all states in the root vertex. In fact, the transport properties remain qualitatively the same. In this setting the trapping probability is just $1/(2L+1)$ of the trapping for the graph in FIG. \ref{fig:multiple_loops} thanks to stretching of the trapped states on the branches. Explicitly, the average ATP is
\begin{align}
\label{star}
\overline{q} = 1 - \sum_{k=1}^{n-2}\frac{k\cdot 1+1\cdot k^2}{N_k^2}\frac{1}{n} = 1 - \frac{1}{2L+1}\left(1-\frac{2}{n}\right).
\end{align}
In fact, we could even consider $L=0$, which brings us directly to the graph with just loops in the initial vertex shown FIG. \ref{fig:multiple_loops}, up to a sink issue: Clearly, reducing the sink branch to length 0 and making the root vertex to be the sink vertex would result in complete transport of any initial state. We just have to keep the sink chain of length at least 1.\footnote{We know that it actually does not matter how the sink branch looks like as long as it does not contain any non-sink unpaired loops.} Then the equation (\ref{star}) also holds for $L=0$ and we recover (\ref{loops_exampe}). We see that adding branches still decreases the ATP. This can no more be reasoned beforehand as pure addition since the initial subspace is modified. Nevertheless, it can be expected as the influence of the absorbing branch with the sink is being diluted by the increasing number of branches inducing trapping.

\subsection{Geometry modified by insertion}

Importantly, the process of pure addition must be distinguished from the second type of graph modification which is \textit{insertion} of elements into a graph. In this case we add elements that modify the previous structure of the graph. The insertion is realised by breaking an existing edge, adding a new vertex between the previously adjacent vertices and connecting the two original vertices with the new one by two new edges. It is motivated by the previously studied lengthening of a ladder graph presented 
in \citep{ladder_cayley,etsuo_ladder}. It was shown that lengthening a ladder graph (terminated with unpaired loops at the ends) surprisingly increases the ATP for transport from an initial vertex at one end of the ladder to a sink vertex at the other end. The effect was shown to be caused by "stretching" of a C-type trapped state. Due to the insertion of vertices and edges between the unpaired loops at the ends of the ladder, a C-type trapped state connecting these loops gains more elements and due to normalization looses overlap with the initial subspace. More A-type states are added with the extension of the ladder, but their gradual contributions decrease exponentially, while the overlap of the connecting C-type state with the initial subspace decreases polynomially. This also shows that adding cycles outside of the initial subspace is substantially less effective in decreasing the ATP then adding unpaired loops into the initial subspace.

\begin{figure}
  \includegraphics[width=0.45\textwidth]{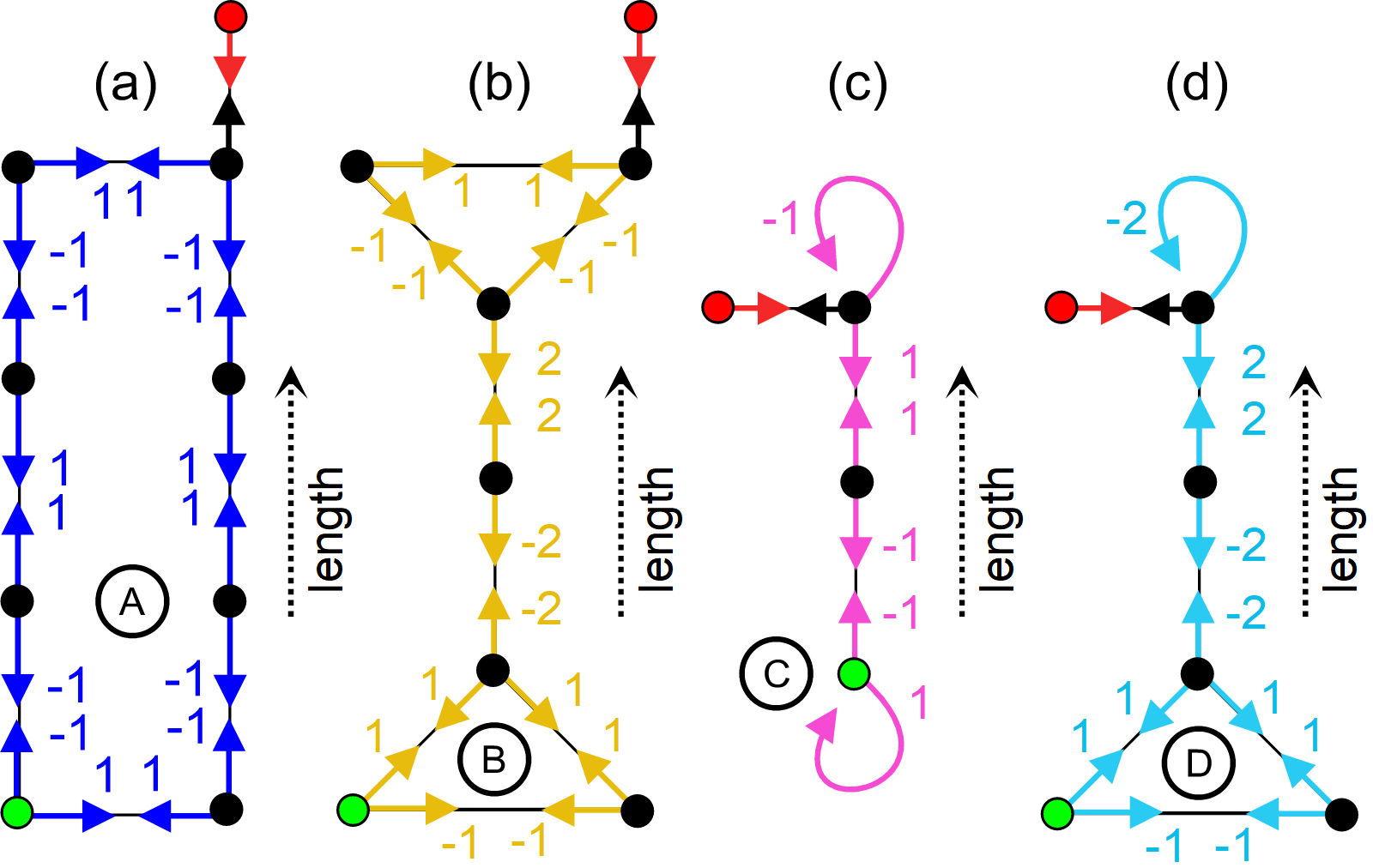}
  \caption{Examples of minimalistic structure graphs supporting the presence of connecting trapped states: (a) a graph with an A-type state, (b) a graph with a B-type state, (c) a graph with a C-type state and (d) a graph with a D-type state. The red vertex indicates the sink and the green vertex indicates the initial subspace.} 
  \label{fig:connecting_states}
\end{figure}

Let us now discuss in more detail which graph geometries tend to exhibit this effect. We have actually seen the same effect of ATP increase by extension of a structure by insertion for the star-like graph above -- the ATP increases with the length of branches $L$. For both the ladder and the star-like graph some C-type states terminated by loops are being stretched. For the ladder without loops at the ends the ATP actually just decreases with length. One might get the impression that the counter intuitive effect of the ATP increase is conditioned by loops added as termination elements. This is not true since in fact the effect can be caused by trapped states of any type. This is best seen if we construct structure graphs so that they only allow a single sr-trapped state. The simplest graphs for each of these types are depicted in FIG. \ref{fig:connecting_states}. In all of them the corresponding unique sr-trapped state is stretched by inserting vertices and edges into the graph. 

As there is only one sr-trapped state in each of the graphs in FIG. \ref{fig:connecting_states}, there is no need for orthogonalization and the ATP can be calculated directly. We denote the number of edges in the connecting path as $L$. For FIG. \ref{fig:connecting_states} (a) let $L$ denote the number of vertical edges minus one. Therefore, all graphs in FIG. \ref{fig:connecting_states} are of length $L=2$ and the minimal length is 0 in all cases. The average ATP for each case is given just by the corresponding normalization constant $N$ of the trapped state, in particular
\begin{align}
\label{increase_atp_examples}
\overline{q}_a &= 1 - \frac{1}{N_a^2}(1+1)\frac{1}{2} = 1 - \frac{1}{8+4L}, \\ \nonumber
\overline{q}_b &= 1 - \frac{1}{N_b^2}(1+1)\frac{1}{2} = 1 - \frac{1}{12+8L}, \\ \nonumber
\overline{q}_c &= 1 - \frac{1}{N_c^2}(1+1)\frac{1}{2} = 1 - \frac{1}{2+2L}, \\ \nonumber
\overline{q}_d &= 1 - \frac{1}{N_d^2}(1+1)\frac{1}{2} = 1 - \frac{1}{10+8L} 
\end{align}
for graphs in FIG. \ref{fig:connecting_states} (a), (b), (c), and (d) respectively. Note that using this expression  for $\overline{q}_c$ with $L=0$ means that we do include both loops into the initial subspace and do not include the edge going to the sink despite it belonging to the same vertex subspace. 

While for example for $\overline{q}_b$ the increase of the ATP with length has a quite low magnitude -- it goes from $11/12$ for $L=0$ and asymptotically towards $1$ for $L\rightarrow \infty$, it is much more prominent for $\overline{q}_c$ ranging from $1/2$ to $1$. This is caused by higher number of vector elements on the termination cycles in (b) compared to the loops in the case (c).

The example graphs have been chosen to demonstrate, in the most striking way, the discussed transport properties. They are minimalistic in their strućture but as crucial building block could be found in, at first sight, more complicated structures. The case of a more complex structure follows in the section on prism graphs.

\subsection{Example: Prism graphs}

\begin{figure}
  \includegraphics[width=0.3\textwidth]{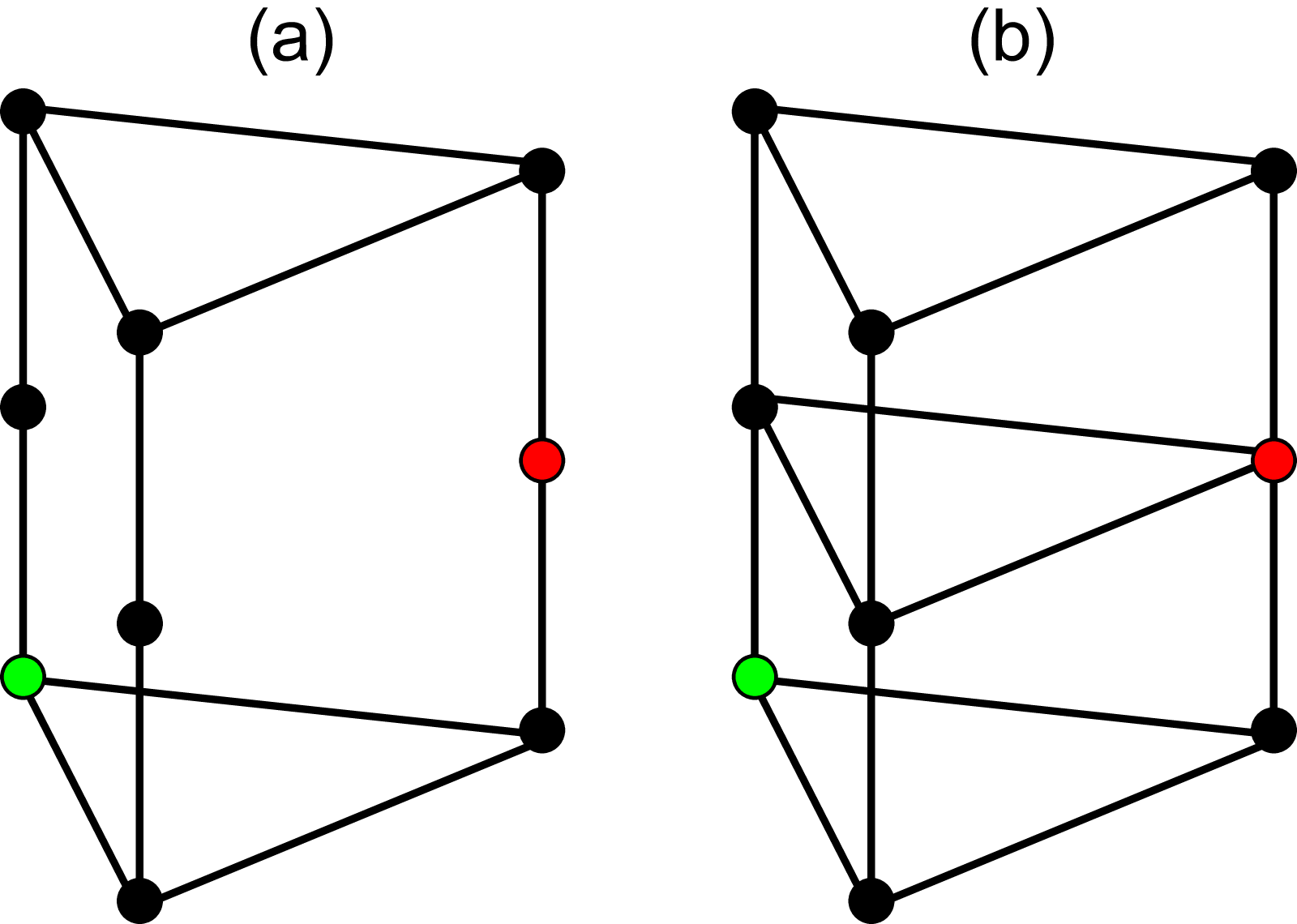}
  \caption{Examples of (a) a hollow prism and (b) a stacked prism, both triangular ($n=3$) and of height $H=2$. Green color indicates the initial vertex and red color indicates the sink.}
  \label{fig:hollow_stacked}
\end{figure}

We consider two versions of graphs, which can be viewed as a prism. First, we consider a "hollow prism" (FIG. \ref{fig:hollow_stacked} (a)) which is composed of two $n$-cycles of the same length (the bases) connected by $n$ chains with $H-1$ vertices and $H$ edges. Here $H$ represents the height of the prism and we assume $H\geq 2$. Second, we consider "stacked prism" (FIG. \ref{fig:hollow_stacked} (b)) which is similar, but also contains edges forming the $H-1$ inner cycles. One of the vertices not belonging to any of bases is chosen to act as a sink. The initial subspace is in one vertex in the base. For simplicity, let us fix a spatial representation of the graphs in vertical position with the base with the initial vertex at the bottom and the other base at the top.

\begin{figure}
  \includegraphics[width=0.5\textwidth]{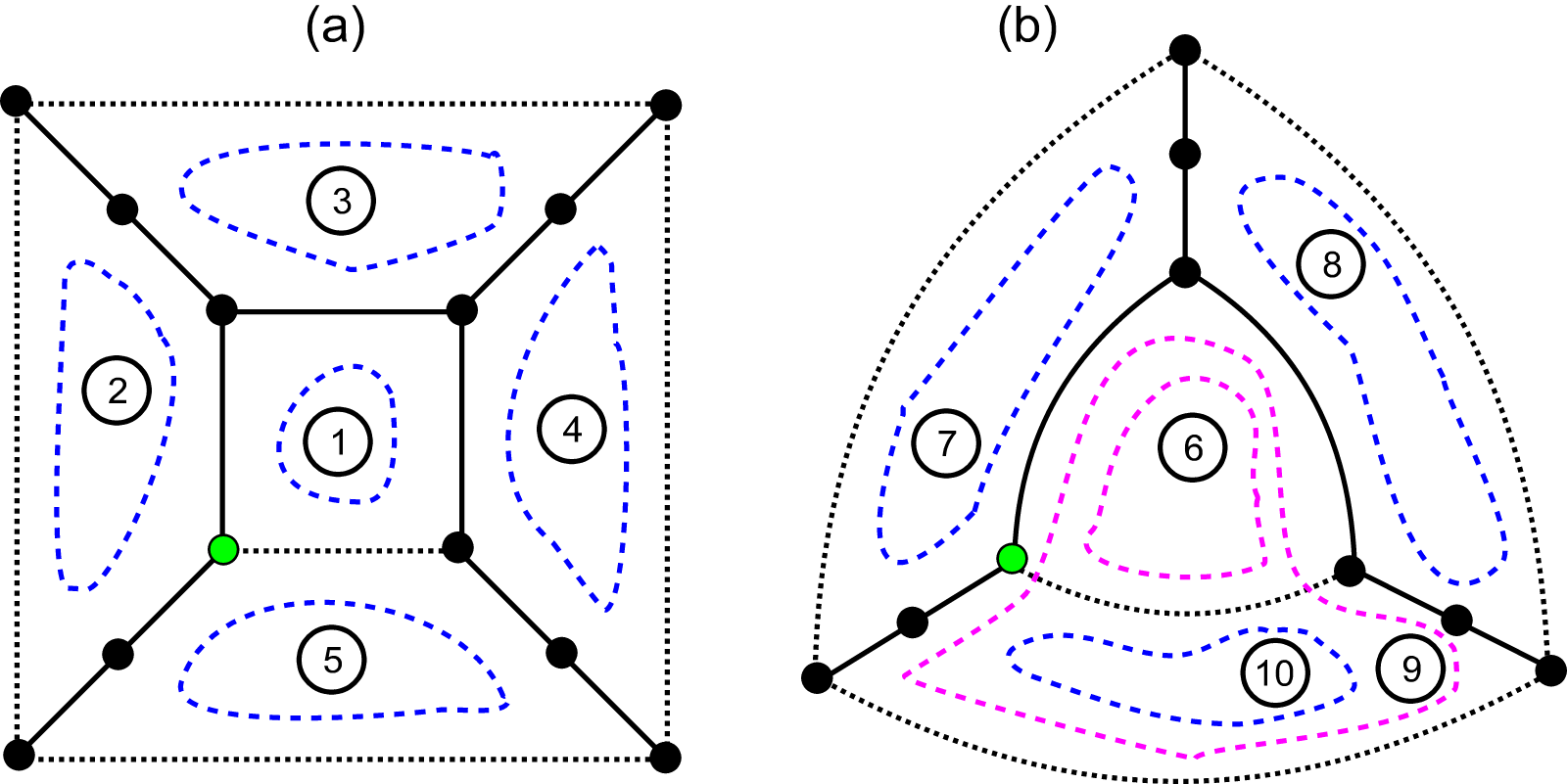}
  \caption{Depiction of a hollow prism with (a) $n=4$ and (b) $n=3$, both with $H=2$ and without sink, in a planar representation. The small square/triangle represents the bottom base and the big one represents the top base. Green color indicates the initial vertex. Dotted edges are not present in the spanning tree and, therefore, each corresponds to one fundamental cycle depicted by dashed line cycles. Blue dashed lines indicate even cycles with a directly corresponding A-type states and magenta lines indicate odd cycles.}
  \label{fig:hollow_no_sink}
\end{figure}

\textbf{Hollow prism:} Let us start with the hollow prism and for now without the sink. The prism has $n(H+1)$ vertices and $2n+nH$ edges, resulting in the need of $n+1$ states in the basis for even $n$ (bipartite graph) or $n$ states for odd $n$. We can create the spanning tree by removing all edges of the top base and one edge in the bottom base as seen in FIG. \ref{fig:hollow_no_sink} (a). For even $n$ we first form the A-type state on the bottom edge (number 1 in the figure). Another $n-1$ A-type states on the vertical faces (numbers 2, 3, 4) can be constructed easily. For the last one, we can use the edge recovered in the first state and make a state on the single face just as all the others (number 5). As was noted before, we are just using a linear combination of the state obtained directly from the corresponding fundamental cycle and the first constructed state (number 1). When $n$ is odd (FIG. \ref{fig:hollow_no_sink} (b)), there is an odd fundamental cycle on the bottom face (number 6) which cannot be associated with an A-type state.
We form the $n-1$ non-problematic A-type states on vertical faces (numbers 7, 8). Lastly, we add the final odd cycle encircling one vertical face and the bottom base (number 9). By joining the two highly overlapping odd cycles (number 6 and 9) we obtain the last state (number 10). It is constructed as a B-type state by connecting two odd cycles but at the end it is just an A-type state on the remaining vertical face.\footnote{Note that we can alternatively replace one of the A-type states by a linear combination of all $n$ of them and obtain a B-type state connecting the two odd cycles on the bases.} 

\begin{figure}
  \includegraphics[width=0.5\textwidth]{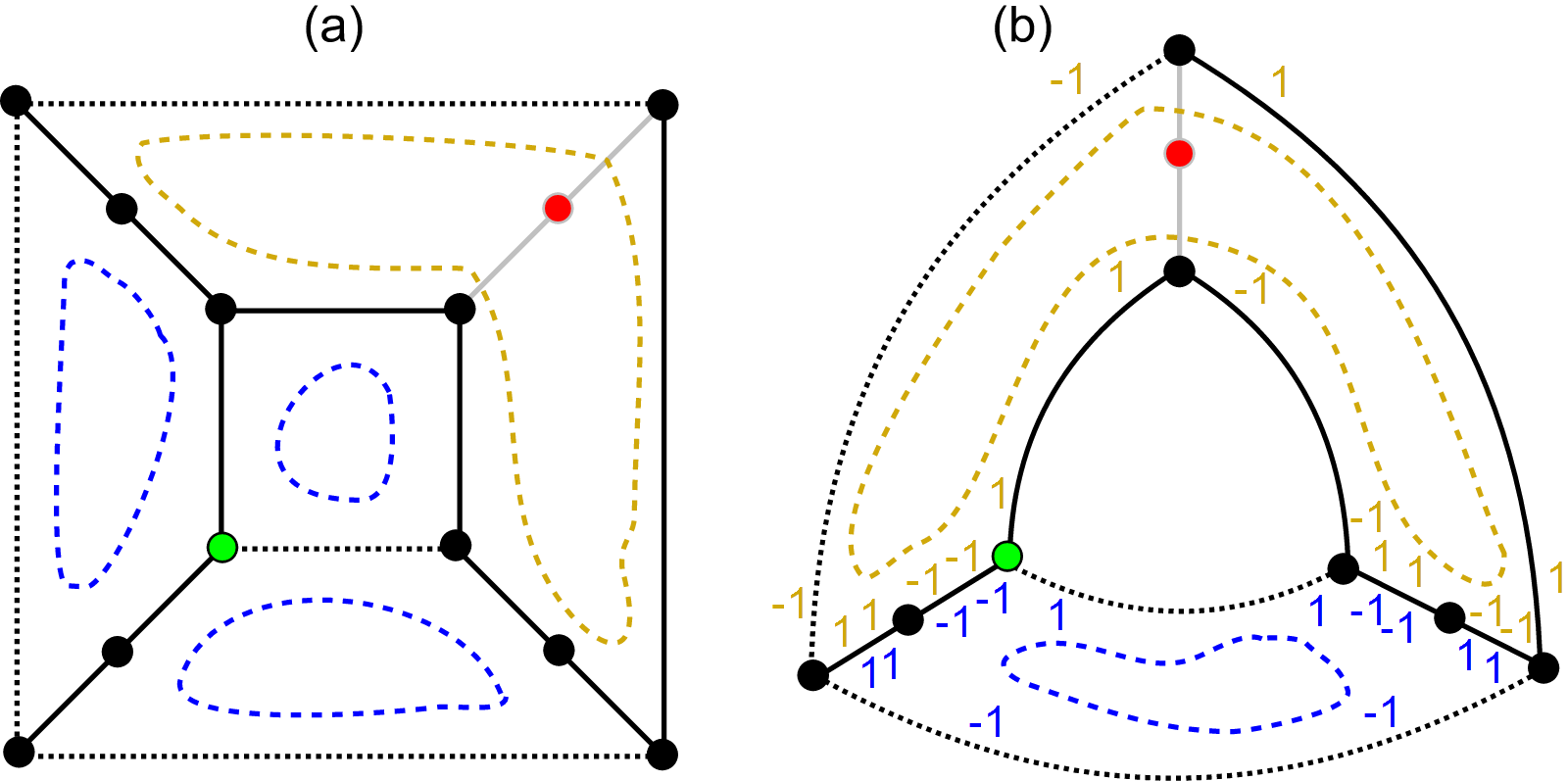}
  \caption{Depiction of a hollow prism with (a) $n=4$ and (b) $n=3$, both with $H=2$ and with sink, in a planar representation. The small square/triangle represents the bottom base and the big one represents the top base. Green color indicates the initial vertex and red color indicates the sink. Gray edges are not present in the reduced graph $\mathcal{G}_0$. Dotted edges are not present in the spanning tree and, therefore, each corresponds to one fundamental cycle depicted by dashed line cycles. Blue dashed lines indicate even cycles with a directly corresponding A-type states present in the graph without sink and orange lines indicate even cycles and the corresponding A-type trapped states modified by the presence of the sink. For the triangular prism also the elements of the two trapped states are presented.}
  \label{fig:hollow_sink}
\end{figure}

Now consider the sink on one chain of vertices ("edge" of the prism) as seen in FIG. \ref{fig:hollow_sink}. We choose the spanning tree so that it contains both edges on the bottom base that are adjacent to the chain with the sink. By considering $\mathcal{G}_0$ with the sink vertex and adjacent edges removed, we must recover one edge from the top base into the spanning tree. Now the whole construction of the basis of trapped states is the same as without the sink, but only one A-type state gets formed on the two vertical faces of the prism that share the sink vertex. Overall, we have $n-1$ sr-trapped states on the vertical faces and one additional sr-trapped state on the bottom base if $n$ is even. Hence, for the hollow prism, increasing the height $H$ does not create new sr-trapped states. However, the height determines the length of the even cycles corresponding to the sr-trapped states on vertical faces, which affects their normalization. For the state corresponding to the cycle going around the sink the normalization is $N'_H = 2\sqrt{H+2}$, and for the remaining $n-2$ states it is equal to $N_H = 2\sqrt{H+1}$ . Our basis is not orthogonal. Nevertheless, any orthogonal basis will be formed by linear combinations of these states and with growing $H$ the trapping will be decreasing approximately as $1/H$. (For even $n$ there is also the non-decreasing contribution to trapping given by the state on the bottom base.) For the simplest case of a triangular prism ($n=3$), the only two sr-trapped states are actually orthogonal and so we can easily calculate the average transport probability as
\begin{align*}
\overline{q} &= 1 - \frac{1}{4+4H}(1+1)\frac{1}{3} - \frac{1}{8+4H}(1+1)\frac{1}{3} = \\
&= 1 - \frac{1}{6}\left(\frac{1}{H+1}+\frac{1}{H+2}\right).
\end{align*}

We see that the counter-intuitive effect of transport increasing with the length of the structure can be present even in a situation where there are no unpaired loops and the sink is placed directly in the main structure and not as an additional vertex sticking out of it. 

\textbf{Stacked prism:} For the stacked prism without sink we can actually use the very same spanning tree as for the hollow prism -- all the horizontal connecting links are removed except $n-1$ edges in the bottom base and all the vertical links are kept. Now the construction can be done in a similar way going from the bottom face upwards creating $n\cdot H$ A-type states on the vertical faces and one more A-type state on the bottom base for even $n$. With the sink, 4 edges are removed completely when transitioning from $\mathcal G$ to ${\mathcal G}_0$, two of which belonged to the spanning tree. One more edge has to be kept in the spanning tree for its connectedness. Overall, four A-type states on the faces adjacent to the sink vertex are replaced by a single A-type state encircling the sink.    

While for the hollow prism the extension in height can only be achieved by the process of insertion, extending a stacked prism above the sink can be thought of as pure addition. Therefore, the ATP can only decrease in this situation as new trapped states are simply added and no existing are modified. In contrast, extension below the sink must be considered as insertion between the sink and the initial vertex. Here, for the prediction of transport properties, we utilise the fact that the contribution of trapped states decreases with their increasing distance from the initial position. This is clear as the limit trapping for an infinite structure is bounded by 1.\footnote{It was actually shown numerically for the ladder in \cite{ladder_cayley} and carbon nanotube structures in \cite{nanotubes}, and actually confirmed analytically for the ladder graph in \cite{etsuo_ladder} that the contribution drops exponentially fast in these cases.} Inserting a layer between the sink an the initial subspace therefore results in more trapping as the states with lower contribution are removed by the sink. Therefore, the ATP decreases with the extension of a stacked prism regardless of whether we insert layers below the sink or add layers above the sink.

Let us note a distinction between the hollow prism and the stacked prism. When extending a hollow prism the number of sr-trapped states remains constant, but the individual states are modified - they cover a longer cycle. In contrast, for the stacked prism the number of sr-trapped states increases with the extension of the prism, however, the states do not change. With the exception of the state surrounding the sink they can be constructed on 4-cycles in the structure graph. This results in fundamentally different transport properties of the hollow and stacked prism.

\subsection{Role of the sink position}
\begin{figure}
  \includegraphics[width=0.4\textwidth]{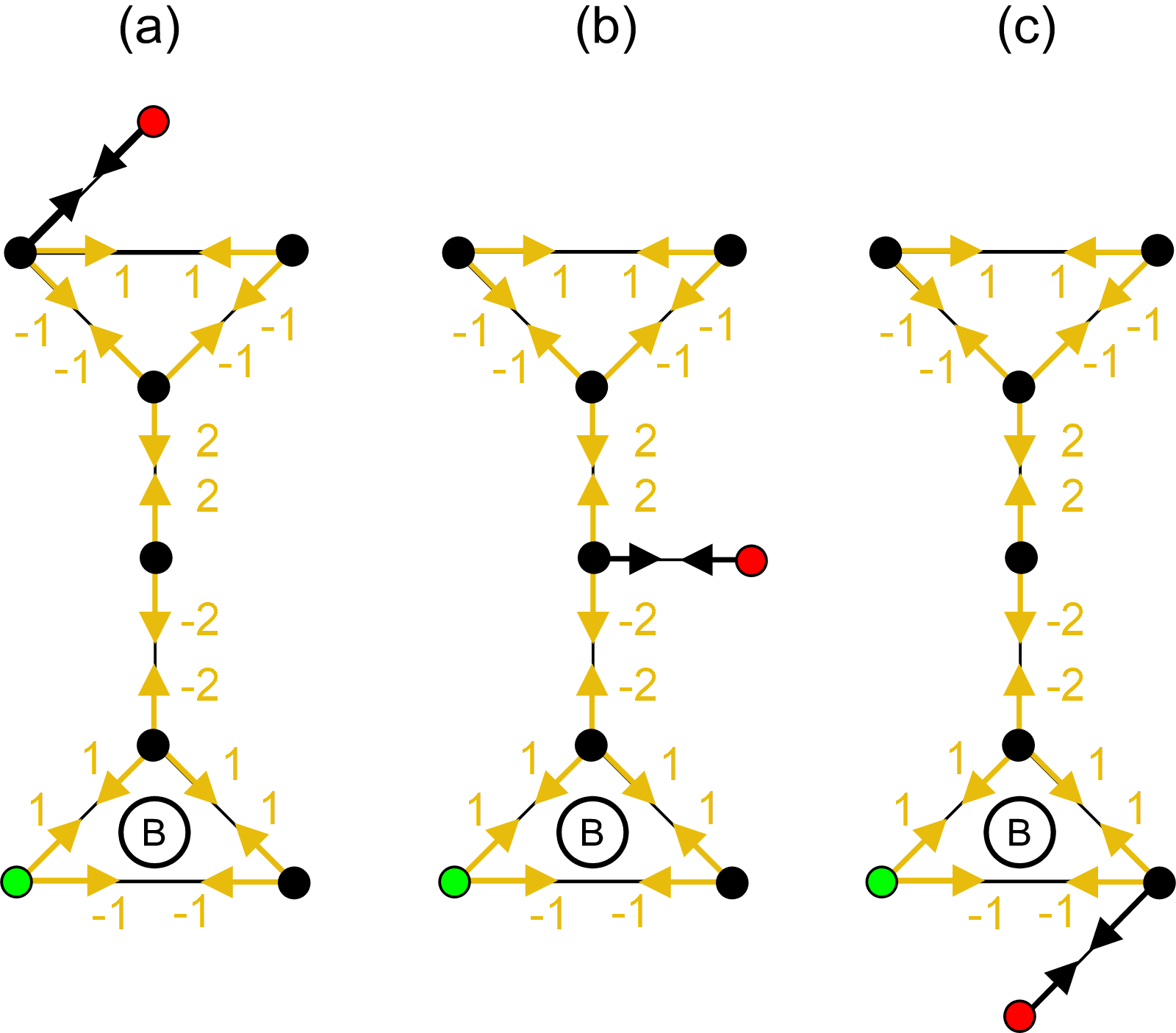}
  \caption{Examples for placement of the sink depicted by a red vertex. The initial subspace is in the green vertex. In all three cases the ATP is the same as there is only one sr-trapped state and it does not overlap with the sink.}
  \label{fig:sink_placements}
\end{figure}

Let us finally explore the role of the \textit{sink placement}. The situation described in \citep{ladder_cayley} where the ATP actually grows with the increasing distance separating the initial vertex and the sink is very counter intuitive. It becomes more understandable when we realize, that there are in fact two independent processes. The increase of the ATP is caused by stretching of the C-type connecting state. Simultaneously, the sink is being moved away from the initial point. Nevertheless, the only thing that matters is that the sink does not remove the connecting sr-trapped state. It can, therefore, be placed anywhere in the graph and its distance from the initial point is irrelevant for the investigated effect of the increase of ATP. FIG. \ref{fig:sink_placements} shows a simplified situation of graphs with a single B-type sr-trapped state with varying position of the sink. ATP is the same in all three cases, but the distance of the sink vertex (red) and the initial vertices (green) differs and also changes differently with extension of the graph. 

In the prism example, the position of the sink on a chain of vertices of a hollow prism is irrelevant for the ATP as long as the sink is not placed to any of the bases. In contrast, the vertical position of the sink in a stacked prism influences the set of sr-trapped states and, therefore, modifies the ATP.

We give one more example related to the placement of the sink. We imagine a network with a single source and multiple potential receivers. In our formalism, we represent the situation as a percolated reflecting Grover quantum walk on a graph with a localised initial subspace and multiple possible sink vertices. In every realization of a quantum walk initiated in the source (representing the transmission of a signal), only one vertex is chosen to act as a sink. The others act just as regular vertices. With our results, we can for example see that if each of the receivers is realized just as a degree-one vertex attached to the graph as in FIG. \ref{fig:sink_placements}, the choice of the receiver has no influence on the set of sr-trapped states. Therefore, the probability of an asymptotic transfer is the same regardless of the receiver chosen. This holds for all initial states. The initial state does influence the transfer probability, but always in the same way. In contrast, would the receivers be realized as vertices of higher degrees included in fundamental cycles of the graph or as vertices with unpaired loops or would the non-active receivers be replaced by unpaired loops instead of staying as degree-one vertices, the choice of the receiver would influence the ATP through the change of the set of sr-trapped states.

\section{Conclusions}
\label{conclusions}

Quantum walks, as generic models of quantum transport, allow for several peculiarities not seen in classical walks. Among such is trapping found for a number of walk "geometries" represented by a corresponding graphs defining the essentials of the walkers motion. In this work we address the phenomenon of asymptotic
trapping in coined flip-flop Grover quantum walks, which
allows  the walker to indefinitely evade a sink placed in
the underlying graph with non-zero probability. For the
version of a quantum walk disrupted by dynamical percolation of edges we prove that the asymptotic trapping is given solely by trapped eigenstates of a walk and provide a complete and general recipe for the construction of their basis for a walk on arbitrary finite simple connected graph. Supplemented with (numerical, when other methods are not available) orthogonalization, this allows to determine the asymptotic transport probability for arbitrary initial states of the walk. Using these results for a very general class of graphs (in contrast to previous restriction to planar 3-regular graphs) we reveal the principal properties of graphs determining asymptotic transport properties of induced percolated flip-flop Grover quantum walks. We also show how changes of additional structure parameters of the graphs (e.g. the height of a prism) influence the asymptotic transport.

There are several important points to be listed. We show the crucial importance of the initial state, where some initial states can lead to complete transport while others to significant trapping. We note that the walker can only be trapped by states directly overlapping (after orthogonalization) with the initial state and the walker can not be trapped along the way. We clarify the difference between modification of a graph by pure addition and insertion of elements. While pure addition can only decrease transport, insertion can actually lead to transport improvement despite increasing the distance of the initial position and the observed terminal position represented by a sink. We discuss the influence of sink placement, where only the overlap with trapped states matters and e.g. the distance between the sink and the initial state is irrelevant for the asymptotic transport probability. We demonstrate the above arguments on examples of graphs accommodating a single trapped state, a star-like graph and graphs of a hollow prism and stacked prism.

In a number of the investigated examples we derived explicit analytical forms of averaged ATP. In all these cases we found, that if the average ATP increases with the size of the system this growth is roughly proportional to the inverse size of the system. These cases confirm that whenever trapped eigenstates are modified by extension of the graph but no new are added the ATP may only increase. However, in the general situation new trapped states can be created. These are two effects which compete against each other and the concrete change of the ATP depends significantly on the details of the arrangement.

Our results will be certainly useful for studies of transport phenomena on quite general graphs. The possibility to trap or release an excitation -- a walker -- to the sink is of relevance to several problems of solid state physics especially the propagation of excitation along macromolecular fibers with side-branches. However we should keep in mind that our studies assume the possibility to manipulate, design or engineer the corresponding underlying graph.    

{\it Acknowledgements:} The authors acknowledge the financial support from RVO14000 and ”Centre for Advanced Applied Sciences”, Registry No. CZ.02.1.01/0.0/0.0/16 019/0000778, supported by the Operational Programme Research, Development and Education, co-financed by the European Structural and Investment Funds.

\appendix
\section{Exclusion of mixed states in the asymptotic subspace}
\label{appendix_non_p}

We prove that the asymptotic subspace of the PCQW with the flip-flop shift operator and the Grover coin on a general structure graph is spanned by pure states and one additional mixed state proportional to the identity operator. Here we do not consider sinks in the graph. (For a quantum walk with sink we need to subsequently find its subspace orthogonal to the sink subspace.) We follow closely the procedure used in \citep{theory}.

The time evolution of our system is given by a so called random unitary operation as (\ref{timeevolutionpercolated}). The starting point of our search for the asymptotic subspace is the attractor equation (\ref{attractors_main}), which we recapitulate here for convenience
\begin{equation}
U_K X_\lambda U_K^\dagger = \lambda X_\lambda\,, ~~{\rm for~all}~K\in 2^\mathcal{E},
\label{attractors}
\end{equation}
where $|\lambda|=1$. An approach tailored for the case of quantum walks was given in \cite{asymptotic1}\footnote{Note that having chosen the reflecting shift operator we do need to worry about the order of the shift operator and the coin operator in $U$ as discussed in \cite{theory}}, where the splitting of the evolution operator into the shift operation and the coin operation is used to decompose the solution into the coin condition
\begin{align}
\label{coin_cond}
C X C^\dagger &= \lambda X
\end{align}
and the shift condition 
\begin{equation}
\label{shift_cond}
R_K X R_K^\dagger = R_L X R_L^\dagger\,, ~~{\rm for~all}~K,L\in 2^\mathcal{E}.
\end{equation}
The shift condition can be cast into an element-wise form \citep{theory}. Let $e\in E$ and $f\in E$ be directed edges in the state graph and tilde represents the other edge in a pair sharing the same undirected support edge. (The tilde operation maps the unpaired loops to themselves.) We denote the matrix elements of the attractor as $X^{e}_{f} = \langle e|X|f\rangle$. For $f\neq e$ and $f\neq \tilde{e}$ we obtain the shift condition
\begin{align}
\label{full_shift_cond}
X^{e}_{f} &= X^{\tilde{e}}_{f} = X^{e}_{\tilde{f}} = X^{\tilde{e}}_{\tilde{f}},
\end{align}
for $f=e$ the shift condition reduces to
\begin{align*}
X^{e}_{e} = X^{\tilde{e}}_{\tilde{e}}
\end{align*}
and for $f = \tilde{e}$ to
\begin{align*}
X^{e}_{\tilde{e}} = X^{\tilde{e}}_{e}.
\end{align*}
The shift condition originates from the percolation and the relaxation of the shift condition for cases $f=e$ and $f = \tilde{e}$ originates from the fact that a single undirrected support edge can not be open and closed simultaneously in one configuration of the percolation graph.

As introduced in the main text we utilise the concept from \cite{asymptotic2} where p-attractors are constructed by (\ref{p_attractors_construction}) from the common eigenstates -- solutions of the equation (\ref{p_attractors_equation_main}), which we also repeat here for convenience
\begin{equation}
\label{p_attractors_equation}
U_K \ket{\phi} = \alpha \ket{\phi}\,, ~~{\rm for~all}~K\subset 2^\mathcal{E}.
\end{equation}
Also the set of equations (\ref{p_attractors_equation}) can be split into the coin condition 
\begin{align}
\label{p_one}
C \ket{\phi} &= \alpha \ket{\phi}
\end{align}
and the shift condition
\begin{align}
\label{p_shift_cond}
R_K \ket{\phi} = R_L \ket{\phi} \,, ~~{\rm for~all}~K,L\in 2^\mathcal{E}.
\end{align}
The shift condition for common eigenstates can be written element-wise in the vector form as
\begin{align}
\label{p_shift}
\phi_{e}=\phi_{\tilde{e}}\,, ~~{\rm for~all}~e\in E,
\end{align}
where $\phi_{e}$ is the element of $\ket{\phi}$ corresponding to the directed edge $e$. It simply requires the elements for the two edges in a pair to be equal. Importantly, it is a stronger condition than the one for general attractors and in the attractor form it is just (\ref{full_shift_cond}) even in cases $f=e$ and $f = \tilde{e}$. Therefore, the equality 
\begin{align}
\label{broken_shift_cond}
X^{e}_{e} = X^{e}_{\tilde{e}}
\end{align}
holds for all directed edges $e$ in the state graph for every p-attractor, but it is not required for general attractors. We use this fact to show that for the type of quantum walk of our interest the whole asymptotic subspace is spanned by p-attractors and the trivial non-p-attractor proportional to the identity operator.

The coin condition (\ref{coin_cond}) can be split into local conditions in vertices
\begin{align}
\label{coin_equation_locally}
G_m \Xi^u_v G_n^\dagger &= \lambda \Xi^u_v,
\end{align}
where $\Xi^u_v$ represents a $m\times n$ matrix block for a row vertex $u$ (of degree $m$) and column vertex $v$ (of degree $n$) in the whole attractor matrix and $G_m$ and $G_n$ are $m$-dimensional and $n$-dimensional Grover matrices.

The shift condition is then used to combine these blocks into a whole attractor. There are always four matrix blocks, which are bound together by the shift condition if there is an edge between corresponding vertices. For a pair of vertices $u,v\in V$ we denote those as $\Xi^u_u$, $\Xi^u_v$, $\Xi^v_u$ and $\Xi^v_v$. In vertex $u$ the coin is $G_m$ and in $v$ it is $G_n$. If $m\neq n$ the blocks have different sizes and $\Xi^u_v$ and $\Xi^v_u$ are rectangular, not square matrices.
 
The equation (\ref{coin_equation_locally}) can be cast into a standard eigenvalue problem 
\begin{equation*}
(G_m)\otimes(G_n)^\ast \ket{\Xi} = \lambda \ket{\Xi},
\end{equation*}
where the asterisk denotes complex conjugation and $\ket{\Xi}$ represents an attractor block in a vector form, so $\braket{a|\Xi |b} = \braket{a,b|\Xi}$.

\begin{figure}
  \includegraphics[width=0.45\textwidth]{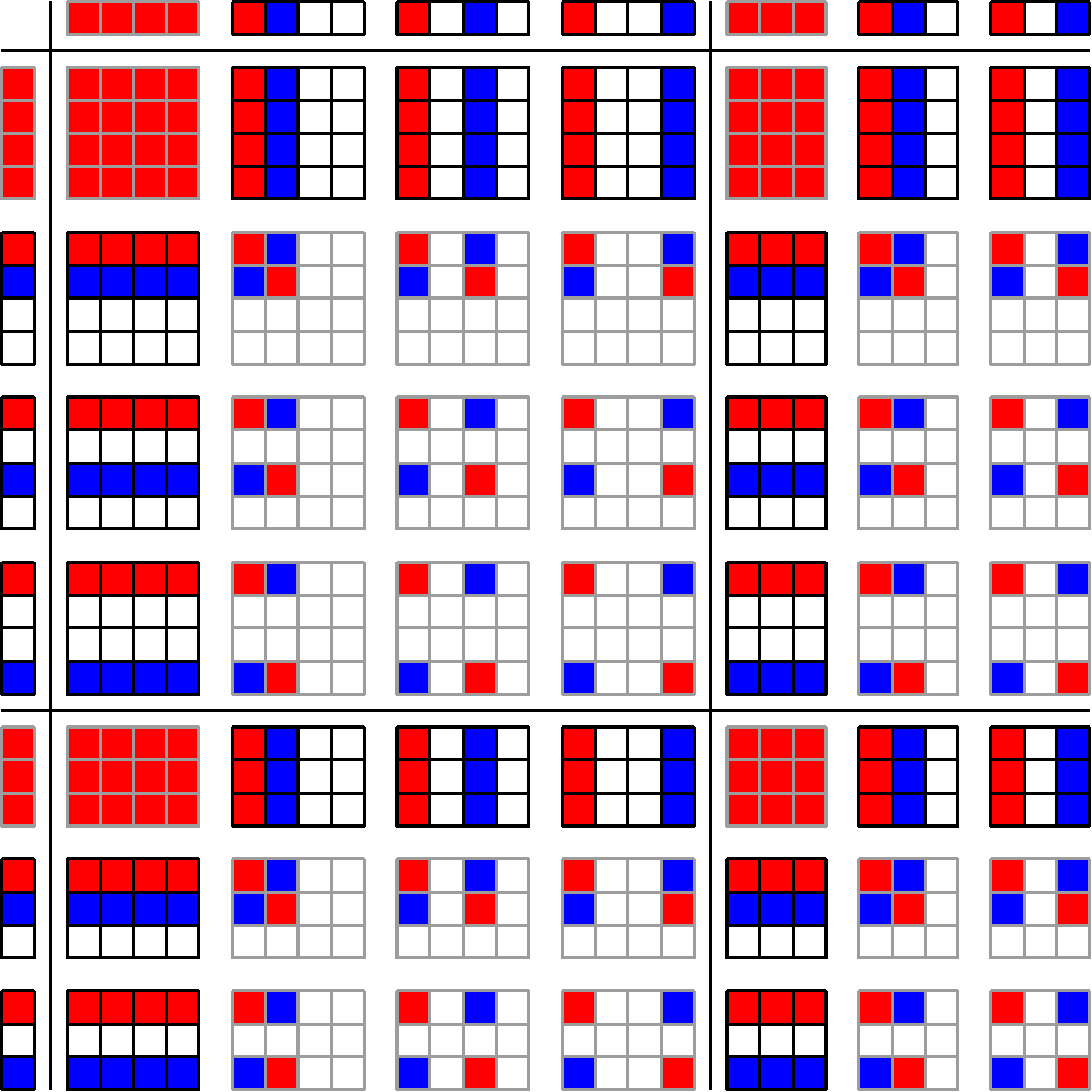}
  \caption{Schematic representation of the basis of attractor blocks for the choice $m=4$ and $n=3$. Red background represents the value 1, blue -1 and white zero. Gray border line indicates correspondence to the eigenvalue 1 and black to the eigenvalue -1. On the top and left are the eigenvectors of $G_4$ and $G_3$. The figure shows blocks for all $\Xi^u_u$, $\Xi^u_v$, $\Xi^v_u$ and $\Xi^v_v$ separated by long lines.}
  \label{fig:tensor_product_construction}
\end{figure}

If $\ket{\psi_1}$ and $\ket{\psi_2}$ are eigenvectors of $G_m$ and $G_n$ respectively, then $\ket{\psi_1}\otimes\ket{\psi_2}^\ast$ is an eigenvector of $(G_m)\otimes(G_n)^\ast$ and therefore $\ket{\psi_1}\bra{\psi_2}$ is a solution of (\ref{coin_equation_locally}). Therefore, we can build the whole basis of solutions of (\ref{coin_equation_locally}) formed by $m\times n$ matrices by combining eigenvectors of $G_m$ and $G_n$, which are described in the main text. The construction is illustrated in FIG. \ref{fig:tensor_product_construction} for the example of $m=4$ and $n=3$.

\begin{figure}
  \includegraphics[width=0.2\textwidth]{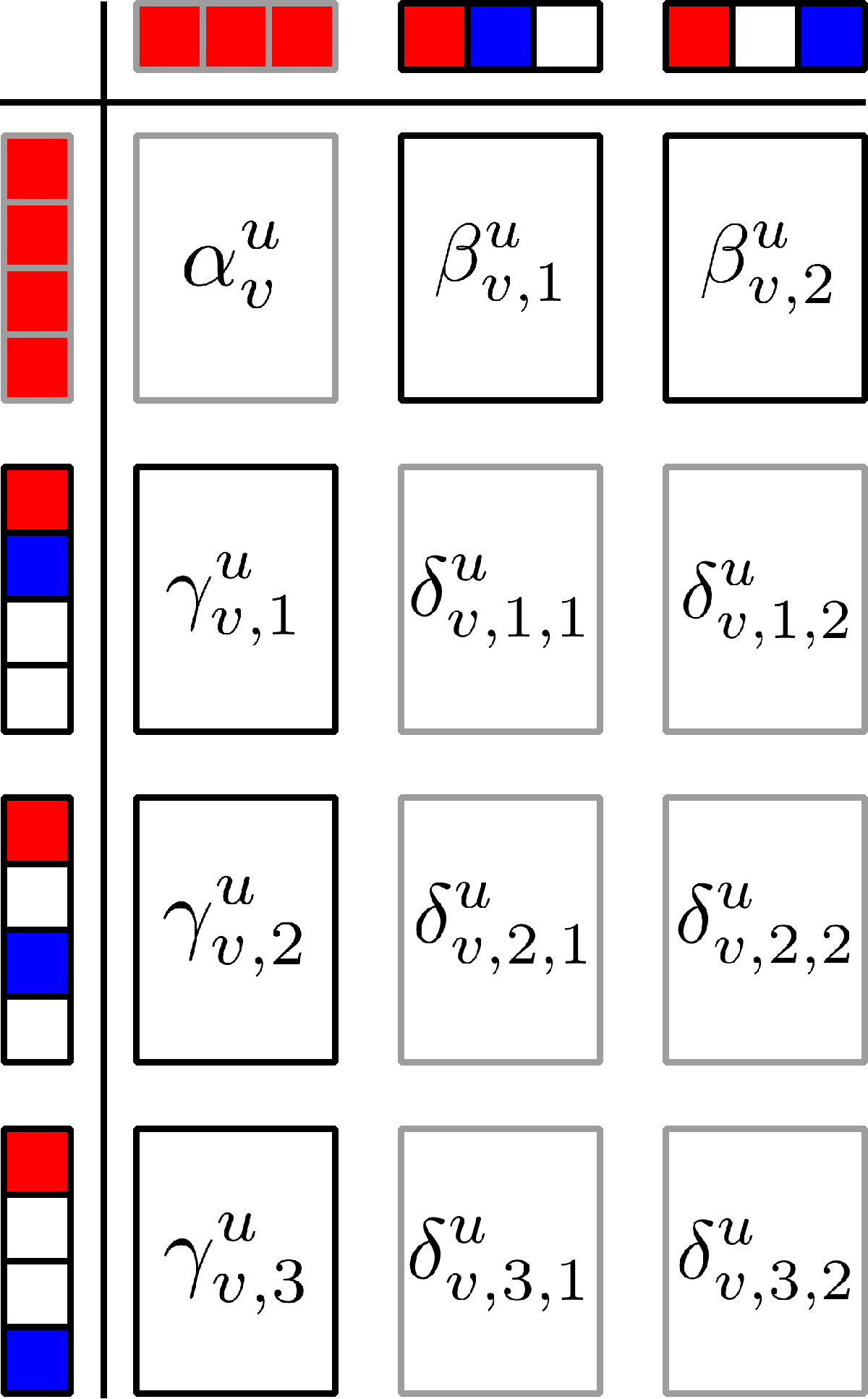}
  \caption{Coefficients of basis blocks shown on the example of basis for the block $\Xi^u_v$ with $d(u)=m=4$ and $d(v)=n=3$.}
  \label{fig:coefficients}
\end{figure}

Let us start with the subspace corresponding to the eigenvalue -1. Here we have $2(m-1)$ base blocks for $\Xi^u_u$, $(m-1)+(n-1)$ base blocks for each of $\Xi^u_v$ and $\Xi^v_u$, and $2(n-1)$ blocks for  $\Xi^v_v$. The final attractor block (e.g. $\Xi^u_u$) is a linear combination the base blocks (the $2(m-1)$ blocks for $\Xi^u_u$). We denote coefficients in this linear combination for the base block as indicated in FIG. \ref{fig:coefficients}. Further, we denote the sums over all admissible indices
\begin{align*}
\sum_j \beta_j = B, \\
\sum_i \gamma_i = \Gamma,
\end{align*}
where the appropriate vertex indices need to be added when used. 

The general form of the block for the eigenvalue -1 (general linear combination of the base blocks) is
\begin{align*}
\Xi=\left[
\begin{array}{cccccc}
B + \Gamma & \Gamma - \beta_1 & \ldots & \Gamma - \beta_j & \ldots & \Gamma - \beta_\nu \\
B - \gamma_1 & - \beta_1 - \gamma_1 & \ldots & - \beta_j - \gamma_1 & \ldots & - \beta_\nu - \gamma_1 \\
\ldots & \ldots & \ldots & \ldots & \ldots & \ldots \\
B - \gamma_i & - \beta_1 - \gamma_i & \ldots & - \beta_j - \gamma_i & \ldots & - \beta_\nu - \gamma_i \\
\ldots & \ldots & \ldots & \ldots & \ldots & \ldots \\
B - \gamma_\mu & - \beta_1 - \gamma_\mu & \ldots & - \beta_j - \gamma_\mu & \ldots & - \beta_\nu - \gamma_\mu
\end{array}
\right]
\end{align*}
where each of $\nu$ and $\mu$ equals either $m-1$ or $n-1$ depending on vertex indices, which are omitted in the above expression. (E.g. for $\Xi^u_v$ the top-left element should be written as $B^u_v+\Gamma^u_v$ and $\mu=m-1$ and $\nu=n-1$.) 

\begin{figure}
  \includegraphics[width=0.25\textwidth]{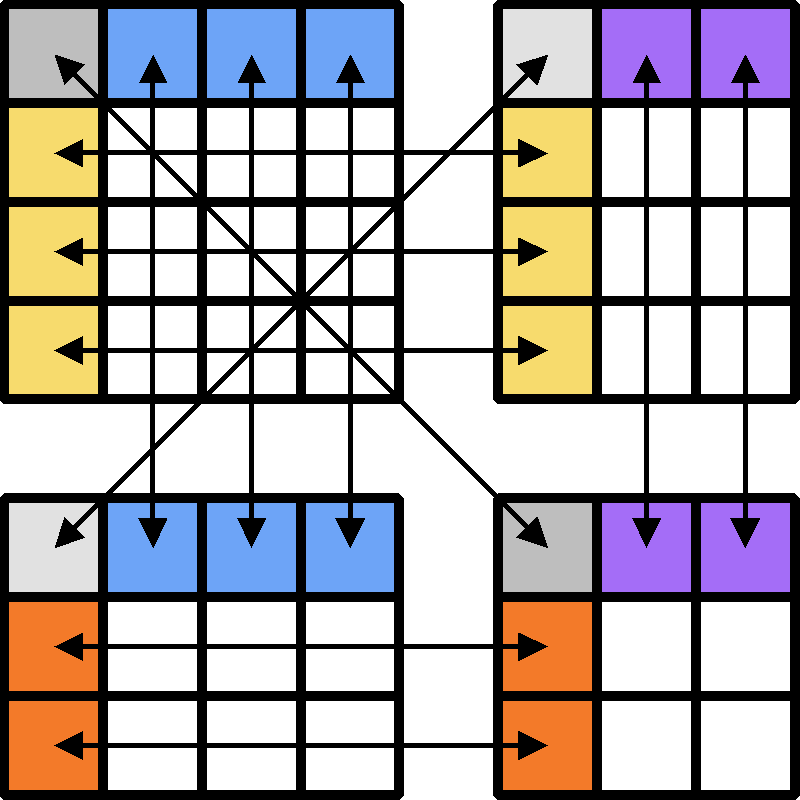}
  \caption{Schematic representation of the shift condition. Arrows represent equalities of elements enforced by the shift condition and colors indicate sets of equalities which are summed together in our method.}
  \label{fig:shift_condition}
\end{figure}

Any attractor must simultaneously satisfy the shift condition. Let us assume that vertices $u$ and $v$ are connected by an edge and we denote the direction of this edge in both of these vertices as $d_1$. (Shuffling labels of the base states has no effect, since the action of the Grover coin is independent of the labeling.) Then the shift condition for corresponding matrix elements of the possible attractor $X$ yields
\begin{align}
\label{shift_cond_2_1}
X^{u d_1}_{u d_1}=X^{v d_1}_{v d_1},\quad X^{u d_1}_{v d_1}=X^{v d_1}_{u d_1}
\end{align}
and
\begin{align}
\label{shift_cond_2_2}
X^{u d_1}_{u d_i}=X^{v d_1}_{u d_i}, \quad X^{u d_1}_{v d_i}&=X^{v d_1}_{v d_i}, \\
X^{u d_i}_{u d_1}=X^{u d_i}_{v d_1}, \quad X^{v d_i}_{u d_1}&=X^{v d_i}_{v d_1}, \nonumber 
\end{align}
where $d_i$ represents arbitrary direction $d_i \neq d_1.$ The shift condition is illustrated in FIG. \ref{fig:shift_condition}.

The condition (\ref{shift_cond_2_1}) can be written in the form of coefficients as 

\begin{align}
\label{conditions_diagonal_m1}
B^u_u + \Gamma^u_u = B^v_v + \Gamma^v_v, \\ \nonumber
B^u_v + \Gamma^u_v = B^v_u + \Gamma^v_u. 
\end{align}

Next we sum all the shift conditions from (\ref{shift_cond_2_2}) over $d_i \neq d_1$ (horizontally and vertically in the illustration in FIG. \ref{fig:shift_condition}), which results in

\begin{align}
\label{conditions_other_m1}
(m-1)B^u_u - \Gamma^u_u &= (m-1)B^u_v - \Gamma^u_v, \\ \nonumber
(m-1)\Gamma^u_u - B^u_u &= (m-1)\Gamma^v_u - B^v_u, \\ \nonumber
(n-1)B^v_v - \Gamma^v_v &= (n-1)B^v_u - \Gamma^v_u, \\ \nonumber
(n-1)\Gamma^v_v - B^v_v &= (n-1)\Gamma^u_v - B^u_v.
\end{align}

When we multiply the equations (\ref{conditions_diagonal_m1}) by $m(n-2)$ and $-(mn-m-n)$ respectively and the equations in (\ref{conditions_other_m1}) by $n$, $n$, $m$ and $m$ respectively and sum them all we obtain 

\begin{align}
\label{conditions_final_m1}
2(mn-m-n)(B^u_u + \Gamma^u_u) &= 2(mn-m-n)(B^u_v + \Gamma^u_v),
\end{align} 
which means

\begin{align}
\label{shift_cond_not_broken_m1}
X^{u d_1}_{u d_1}=X^{u d_1}_{v d_1},
\end{align}
as $mn-m-n \neq 0$ for all positive integers $m,n$. The equality (\ref{shift_cond_not_broken_m1}) represents (\ref{broken_shift_cond}), which has to be broken for an attractor to be a non-p-attractor. Therefore, for the eigenvalue -1 there are only p-attractors for the investigated walks.

Let us move to the subspace corresponding to the eigenvalue 1. Here we have $1+m^2$ base blocks for $\Xi^u_u$, $1+m\cdot n$ base blocks for each of $\Xi^u_v$ and $\Xi^v_u$, and $1+n^2$ for $\Xi^v_v$. These have their coefficients in the linear combination denoted as $\alpha$ and $\delta_{i,j}$ (also with indices for the vertices), where $i$ and $j$ go from 1 to either $m-1$ or $n-1$ depending on the vertices. We denote the sums over all admissible indices (again dropping the vertex indices) as

\begin{align*}
\sum_{i,j} \delta_{i,j} = \Delta, \\
\sum_{i} \delta_{i,j} = \Delta_{\bullet,j}, \\
\sum_{j} \delta_{i,j} = \Delta_{i,\bullet}.
\end{align*}

\begin{figure}
  \includegraphics[width=0.45\textwidth]{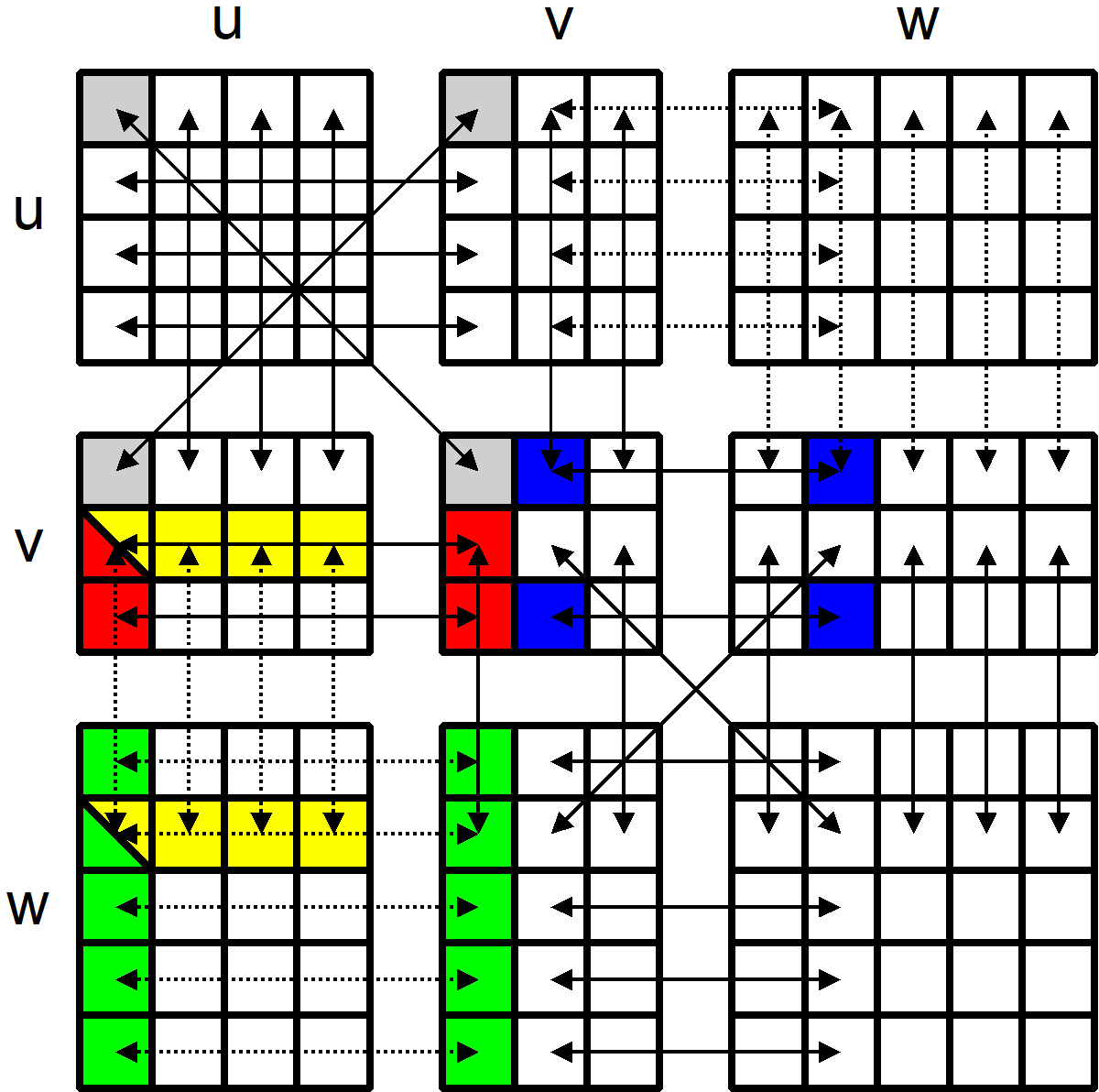}
  \caption{Schematic representation of the shift condition for three vertices. Arrows represent equalities of elements enforced by the shift condition and colors indicate sets of equalities which are summed together. Conditions from (\ref{shift_cond_2_3_p1}) are represented by dotted arrows.}
  \label{fig:shift_condition_3_vertices}
\end{figure}

The general form of the block for the eigenvalue 1 is
\begin{align*}
\Xi=\left[
\begin{array}{cccccc}
\alpha + \Delta & \alpha-\Delta_{\bullet,1} & \ldots & \alpha-\Delta_{\bullet,j} & \ldots & \alpha-\Delta_{\bullet,\nu} \\
\alpha-\delta_{1,\bullet} & \alpha+\delta_{1,1} & \ldots & \alpha+\delta_{1,j} & \ldots & \alpha+\delta_{1,\nu} \\
\ldots & \ldots & \ldots & \ldots & \ldots & \ldots \\
\alpha-\delta_{i,\bullet} & \alpha+\delta_{i,1} & \ldots & \alpha+\delta_{i,j} & \ldots & \alpha+\delta_{i,\nu} \\
\ldots & \ldots & \ldots & \ldots & \ldots & \ldots \\
\alpha-\delta_{\mu,\bullet} & \alpha+\delta_{\mu,1} & \ldots & \alpha+\delta_{\mu,j} & \ldots & \alpha+\delta_{\mu,\nu}
\end{array}
\right].
\end{align*}
Let us assume that vertices $u$ and $v$ are connected by an edge and we denote the direction of this edge in both of these vertices as $d_1$. Assume that $v$ is also connected to $w$ (with subspace of dimension $o$) by an edge and we denote the direction of this edge in both of these vertices as $d_2$. (Again, shuffling labels of the base states has no effect for the Grover coin.) Then the shift condition for the corresponding matrix elements of a possible attractor $X$ requires
\begin{align}
\label{shift_cond_2_1_p1}
X^{u d_1}_{u d_1}=X^{v d_1}_{v d_1},\quad X^{u d_1}_{v d_1}=X^{v d_1}_{u d_1} \\ \nonumber 
X^{v d_2}_{v d_2}=X^{w d_2}_{w d_2},\quad X^{v d_2}_{w d_2}=X^{w d_2}_{v d_2}
\end{align}
and
\begin{align}
\label{shift_cond_2_2_p1}
X^{u d_1}_{u d_i}=X^{v d_1}_{u d_i}, \quad X^{u d_1}_{v d_i}&=X^{v d_1}_{v d_i}, \\ \nonumber 
X^{u d_i}_{u d_1}=X^{u d_i}_{v d_1}, \quad X^{v d_i}_{u d_1}&=X^{v d_i}_{v d_1}, \\ \nonumber 
X^{v d_2}_{v d_j}=X^{w d_2}_{v d_j}, \quad X^{v d_2}_{w d_j}&=X^{w d_2}_{w d_j}, \\ \nonumber 
X^{v d_j}_{v d_2}=X^{v d_j}_{w d_2}, \quad X^{w d_j}_{v d_2}&=X^{w d_j}_{w d_2},   
\end{align}
where $d_i$ and $d_j$ represent arbitrary directions with $d_i \neq d_1$ and $d_j \neq d_2$. Nevertheless, there are further conditions relating elements for $u$ and $w$ directly

\begin{align}
\label{shift_cond_2_3_p1}
X^{u d_1}_{w d_k}=X^{v d_1}_{w d_k} \\ \nonumber 
X^{w d_k}_{u d_1}=X^{w d_k}_{v d_1} \\ \nonumber 
X^{v d_2}_{u d_k}=X^{w d_2}_{u d_k} \\ \nonumber 
X^{u d_k}_{v d_2}=X^{u d_k}_{w d_2},   
\end{align}
where $d_k$ represents arbitrary direction. The shift condition is illustrated in FIG. \ref{fig:shift_condition_3_vertices}.

In this case, we know that there is at least the non-p-attractor proportional to the identity matrix since it trivially fulfills the attractor equation (\ref{attractors}). Therefore, we know that the conditions
\begin{align*}
X^{u d_1}_{u d_1} = X^{u d_1}_{v d_1}
\end{align*}
and
\begin{align*}
X^{v d_2}_{v d_2} = X^{v d_2}_{w d_2}
\end{align*}
representing (\ref{broken_shift_cond}) are not enforced directly by the combination of coin conditions and shift conditions. Instead, we start by proving
\begin{align}
\label{implication}
X^{u d_1}_{u d_1} = X^{u d_1}_{v d_1} \Rightarrow X^{v d_2}_{v d_2} = X^{v d_2}_{w d_2}.
\end{align}
The equalities (\ref{shift_cond_2_1_p1}) can be written as
\begin{align}
\label{conditions_diagonal_p1}
\alpha^u_u + \Delta^u_u &= \alpha^v_v + \Delta^v_v, \\ \nonumber
\alpha^u_v + \Delta^u_v &= \alpha^v_u + \Delta^v_u, \\ \nonumber
\alpha^v_v + \delta^v_{v,1,1} &= \alpha^w_w + \delta^w_{w,1,1}, \\ \nonumber
\alpha^v_w + \delta^v_{w,1,1} &= \alpha^w_v + \delta^w_{v,1,1}.
\end{align}
With the assumption of the implication that we are to prove we have
\begin{align}
\label{conditions_diagonal_p1_with_assumptiion}
\alpha^u_u + \Delta^u_u &= \alpha^v_v + \Delta^v_v = \alpha^u_v + \Delta^u_v = \alpha^v_u + \Delta^v_u.
\end{align}

By summing the equations for red elements in FIG. \ref{fig:shift_condition_3_vertices} and using (\ref{conditions_diagonal_p1_with_assumptiion}) we get
\begin{align}
(n-1)\alpha^v_u - \Delta^v_u &= (n-1)\alpha^v_v  -\Delta^v_v, \\ \nonumber
n\alpha^v_u - (\alpha^v_u+\Delta^v_u) &= n\alpha^v_v - (\alpha^v_v + \Delta^v_v), \\ \nonumber
\alpha^v_u &= \alpha^v_v.
\end{align}
By further summing the equations for green elements we obtain
\begin{align}
o\alpha^w_u &= o\alpha^w_v, \\ \nonumber
\alpha^w_u &= \alpha^w_v
\end{align}
and by summing the equations for yellow elements
\begin{align}
m\alpha^v_u &= m\alpha^w_u, \\ \nonumber
\alpha^v_u &= \alpha^w_u.
\end{align}
Finally, summing the equations for blue elements and using the four previous result allows us to derive
\begin{align}
(n-1)\alpha^v_v - \delta^v_{v,1,1} &= (n-1)\alpha^w_v  -\delta^w_{v,1,1}, \\ \nonumber
n\alpha^v_v - (\alpha^v_v+\delta^v_{v,1,1}) &= n\alpha^w_v - (\alpha^w_v + \delta^w_{v,1,1}), \\ \nonumber
n\alpha^v_v - (\alpha^v_v+\delta^v_{v,1,1}) &= n\alpha^w_u - (\alpha^w_v + \delta^w_{v,1,1}), \\ \nonumber
n\alpha^v_v - (\alpha^v_v+\delta^v_{v,1,1}) &= n\alpha^v_u - (\alpha^w_v + \delta^w_{v,1,1}), \\ \nonumber
\alpha^v_v+\delta^v_{v,1,1} &= \alpha^w_v + \delta^w_{v,1,1},
\end{align}
and therefore after using the last equality from (\ref{shift_cond_2_1_p1}) we have
\begin{align*}
X^{v d_2}_{v d_2} = X^{v d_2}_{w d_2},
\end{align*}
so the implication (\ref{implication}) is proven. Now we just directly apply the reasoning from \citep{theory}. We consider an arbitrary attractor $X$ corresponding to the eigenvalue 1 and choose two vertices $u$ and $v$ connected by an edge in direction $d$. We define a new attractor $Y = X + z I$, with $z = X^{u d}_{v d} - X^{u d}_{u d}$, so $Y^{u d}_{u d} = X^{u d}_{u d} + z = X^{u d}_{v d} + 0 = Y^{u d}_{v d}$. From (\ref{implication}) and the symmetry of the Grover coin it follows that since $Y^{u d}_{u d} = Y^{u d}_{v d}$, it also holds $Y^{u' d'}_{u' d'} = Y^{u' d'}_{v' d'}$ for arbitrary vertices $u'$ and $v'$ connected by an edge in direction $d'$, implying that $Y$ is a p-attractor. Hence, an arbitrary attractor is a linear combination of the trivial non-p-attractor and a p-attractor. We conclude, that there are no other linearly independent non-p-attractors apart from the trivial one proportional to the identity operator.

\end{document}